\documentclass[preprint2]{aastex}

\usepackage{lscape}
\usepackage{graphicx}

\newcommand{\kms}{km~s$^{-1}$~}
\newcommand{\etal}{et al.~}
\newcommand{\HI}{\ion{H}{1}~}
\newcommand{\sol}{\ifmmode_{\mathord\odot} \else $_{\mathord\odot}$ \fi}
\newcommand{\msun}{M$_{\sol}$}

\begin{document}

\title{The Arecibo L-band Feed Array Zone of Avoidance Survey I: \\  
Precursor Observations through the Inner and Outer Galaxy \\
}

\author{
P.A. Henning,\altaffilmark{1}
C.M. Springob,\altaffilmark{2}
R.F. Minchin,\altaffilmark{3}
E. Momjian,\altaffilmark{4}
B. Catinella,\altaffilmark{5}
T. McIntyre,\altaffilmark{1}
F. Day,\altaffilmark{1}
E. Muller,\altaffilmark{6}
B. Koribalski,\altaffilmark{7}
J.L. Rosenberg,\altaffilmark{8}
S. Schneider,\altaffilmark{9}
L. Staveley-Smith,\altaffilmark{10}
W. van Driel\altaffilmark{11}
}

\altaffiltext{1}{Institute for Astrophysics, University of New Mexico, MSC07 4220, 800 Yale Blvd., NE, Albuquerque, NM, 87131, USA}
\altaffiltext{2}{Anglo-Australian Observatory, P.O. Box 296, Epping, NSW 1710, Australia}
\altaffiltext{3}{National Astronomy and Ionosphere Center - Arecibo Observatory, HC3 Box 53995, Arecibo, PR 00612, USA}
\altaffiltext{4}{National Radio Astronomy Observatory, P.O. Box O, Socorro, NM 87801, USA}
\altaffiltext{5}{Max-Planck-Institut f\"ur Astrophysik, Karl-Schwarzschild-Strasse 1, D-85741 Garching, Germany}
\altaffiltext{6}{Department of Astrophysics, Nagoya University, Furo-cho, Chikusa-ku, Nagoya 464-8602, Japan}
\altaffiltext{7}{Australia Telescope National Facility, CSIRO, P.O. Box 76, Epping, NSW 2121, Australia}
\altaffiltext{8}{Department of Physics and Astronomy, George Mason University, Fairfax, VA 22030, USA}
\altaffiltext{9}{Department of Astronomy, University of Massachusetts, Amherst, MA 01003, USA}
\altaffiltext{10}{School of Physics, University of Western Australia, Crawley WA 6009, Australia}
\altaffiltext{11}{GEPI, Observatoire de Paris, CNRS, Universit´{e} Paris Diderot, 5 place Jules Janssen, 92190 Meudon, France}

\begin{abstract}

The Arecibo L-band Feed Array (ALFA) is being used to conduct 
a low-Galactic latitude survey, to map the distribution of 
galaxies and large-scale structures behind the Milky Way through detection 
of galaxies' neutral hydrogen (HI) 21-cm emission.
This Zone of Avoidance (ZOA) survey finds new HI galaxies which lie 
hidden behind 
the Milky Way, and also provides redshifts for partially-obscured galaxies 
known at other wavelengths. 
Before the commencement of the full survey,
two low-latitude precursor regions were observed, 
totalling 138 square degrees, with 72 HI galaxies detected.
Detections through the inner Galaxy generally have no cataloged counterparts in
any other waveband, due to the heavy extinction and stellar confusion.
Detections through the outer Galaxy are more likely to have 2MASS counterparts.
We present the results of these precursor observations, including a catalog of
the detected galaxies, with their HI parameters.
The survey sensitivity is well described by a flux- and linewidth-dependent
signal-to-noise ratio of 6.5. 
ALFA ZOA galaxies which also have HI measurements in the
literature show good agreement between our measurements and previous work.
The inner Galaxy precursor region was chosen to overlap the HI Parkes Zone of
Avoidance Survey so ALFA performance could be quickly assessed.
The outer Galaxy precursor region lies north of the Parkes sky.
Low-latitude large-scale structure in this region is revealed, including an
overdensity of
galaxies near $\ell = 183^{\circ}$ and between 5000 - 6000 \kms in the ZOA.
The full ALFA ZOA survey will be conducted
in two phases:  a shallow survey using the observing
techniques of the precursor observations, and also a deep phase with
much longer integration time, with thousands of galaxies predicted
for the final catalog. 

\end{abstract}


\section{Introduction}

The obscuration due to dust and the high stellar density in our
Galaxy varies from place to place within the Milky Way.
Overall, it blocks $\sim$20\% of the extragalactic Universe at
optical wavelengths and a smaller fraction of the sky at infrared
wavelengths. 
This ``Zone of Avoidance" (ZOA) was recognized even before the
nature of the spiral nebulae was understood.
This sky coverage limitation
does not pose a problem for the study of galaxies themselves,
as there is no reason to believe that the population of obscured galaxies 
should differ from those in optically unobscured regions.
Yet, an accurate knowledge of the mass distribution within our neighborhood
is essential if we are to understand the dynamical evolution of the Local
Group from kinematic studies (e.g., Peebles \etal 2001).
In addition, the discovery of previously unknown nearby galaxies will
further efforts to understand the local velocity field (see Kraan-Korteweg
1986 and Karachentsev \etal 2002).
Mapping more distant hidden galaxies allows us to explore 
the connectivity of large-scale
structure across the Galactic plane.

The ZOA has been successfully narrowed by deep searches in the
optical and infrared, but both fail in regions of high extinction
and stellar confusion.  However, galaxies which contain HI can be found
everywhere, including regions of thickest obscuration, and worst IR confusion.
In the northern hemisphere, the ZOA within $\pm$5 degrees 
of the Galactic plane has been searched at 21 cm, but
only at the high noise level of 40 mJy beam$^{-1}$ (with velocity resolution of
4 \kms), sensitive only to
nearby, massive objects [The Dwingeloo Obscured Galaxies Survey:
43 galaxies uncovered (Henning et al. 1998; Rivers 2000)].
More recently, the HI Parkes
Zone of Avoidance Survey (HIZOA) covered Dec = $-90^{\circ}$ to $+25^{\circ}$ 
at 6 mJy beam$^{-1}$ rms (with velocity resolution of
27 \kms), and detected
about one thousand galaxies (Donley \etal 2005; Henning \etal 2000, Henning \etal 2005
Kraan-Korteweg \etal 2005).

With the installation of the Arecibo L-band Feed Array (ALFA), we now 
have the opportunity to map local large scale structure in HI in the ZOA, 
within the declination limits of the 305-m Arecibo Radio 
Telescope\footnote{The Arecibo Observatory is part of the National Astronomy and
Ionosphere Center, which is operated by Cornell University
under a cooperative agreement with the National Science Foundation.}.
Because many observational targets for pulsar and Galactic HI observers 
are found at low Galactic latitude, we have begun a program of ``commensal''
(meaning simultaneous) observations with Galactic HI and pulsar observers.  
Commensal observing, using multiple backends simultaneously, 
makes efficient use of
observing hours, which is particularly important during the highly oversubscribed
Galactic time.

The project being described here, ``ALFA ZOA'', is the combination of two different
commensal observing projects: (1) a map of the Arecibo sky 
at $\ell=30^{\circ}-75^{\circ}$, $|b|<10^{\circ}$ 
(Inner Galaxy region) with a backend designed for detecting extragalactic HI that
will be used in conjunction with a backend for observing Galactic HI and one
for observing $\sim1.4$ GHz radio continuum emission;  
and (2) a deeper map of the Arecibo sky at $|b|<5^{\circ}$ beginning in the 
Inner Galaxy region with a backend designed to detect extragalactic HI in
conjunction with a 
spectrometer used for observing pulsars.
The extragalactic data from this second survey
will also be used to search for Galactic radio recombination lines.
Both of these projects will reach farther north than HIZOA, and 
will provide higher spatial and velocity resolution than the HIZOA survey.
The second survey will also provide higher sensitivity than HIZOA.

ALFA ZOA is complementary to the three other extragalactic blind surveys that
are currently underway at Arecibo:
(1) the {\it Arecibo Legacy Fast ALFA 
Survey} (ALFALFA, e.g. Giovanelli \etal 2005) which is a large area but relatively
shallow survey, (2) the {\it Arecibo Galaxy 
Environments Survey} (AGES, e.g. Auld \etal 2006) a medium-deep survey, and
(3) the {\it ALFA Ultra Deep Survey} (AUDS, Freudling \etal 2005) a very deep survey
with small sky coverage.

The full ALFA ZOA survey 
has begun, and will take several years to complete.
We present a description of the ALFA ZOA survey and the results of 
ZOA precursor observations that were taken commensally with
two smaller Galactic HI projects.
The first project 
covered 38 square degrees near 
$\ell = 40^\circ$, and the other covered 100 square degrees near $\ell = 192^\circ$.

We offer a description of the ALFA ZOA survey, and then 
present the observational results from the precursor observations.

In \S~2, we motivate the ALFA ZOA survey.  
In \S~3, we describe the early precursor observations.
In \S~4, we describe the data reduction, galaxy recognition and parametrization, 
and selection function.
\S~5 contains an overview of the detected galaxies, their HI parameters and
any counterparts at other wavelengths, and discusses the inner and outer Galaxy results.
\S~6 describes the outlook for the full ALFA ZOA survey. 

\section{The ALFA Zone of Avoidance Survey}

ALFA ZOA is one of several extragalactic HI surveys being undertaken with ALFA.  
The emphasis of this project is to trace the 
local large-scale structure in optically-obscured
regions of the sky, and to better understand the local velocity field.
This project will also make important contributions to studies of the HI properties of galaxies in
different environments, as there are a number of interesting large-scale structures 
which lie at low-Galactic latitude which will be surveyed in HI.
The deep phase of the survey will be deeper
than ALFALFA, and will cover a broader contiguous area of the sky than AGES at similar
depth.
The catalog will be used to construct an HI mass function over a large wedge of space,
and to study the HI properties of galaxies in the various environments within the survey
volume.

\subsection{Large Scale Structure and the Local Velocity Field}

Due to the declination limitations of the Arecibo Radio Telescope, there are two separate regions of the 
ZOA which we can map: the inner Galaxy, $\ell=30^{\circ}-75^{\circ}$, and the outer 
Galaxy, $\ell=170^{\circ}-215^{\circ}$.  
Known galaxies from the Lyon-Meudon Extragalactic Database (LEDA) and HIZOA in both of the 
accessible regions are plotted in the top panels of Figure 1.
Optical surveys are compromised at $A_B \sim 1$ mag, and are very ineffective where 
the extinction $A_B$ reaches 3 mag, as the correspondence of galaxy surface 
density with extinction shows (Fig 1 bottom).  
Further, particularly in the inner Galaxy, the near-infrared Two Micron All Sky Survey 
(2MASS, eg. Skrutskie \etal ~2006), less affected by dust extinction than are optical 
surveys, is still defeated by the high Galactic stellar surface density of the bulge.
Still, we do know something about the large-scale structures we will probe in this 
volume from galaxy distributions above and below the plane. 

In the inner Galaxy ZOA region 
the ZOA intersects the Delphinus void 
(center $\ell, b,$v, $=59^{\circ}, -6^{\circ}, 2500$ \kms; Fairall 1998 used for 
this and all following void locations).  Also lying in the inner Galaxy ZOA is a
portion of a smooth ``sine-wave'' shaped feature that can be traced across the 
whole southern sky (Kraan-Korteweg, Koribalski, \& Juraszek 1999).  
This long, sinuous galaxy overdensity disappears into the ZOA at $\ell\sim 40^{\circ}$.  
The HIZOA survey indicates a possible border 
at $\ell=45^{\circ}$ between the Microscopium void ($\ell, b,$ v $=10^{\circ}, 1^{\circ}$, 
4500 \kms ) and the Cygnus void ($\ell, b,$ v $=67^{\circ}, -9^{\circ}$, 3500 \kms ).  
This area will be covered by the ALFA ZOA full survey.
At high velocities, very little is known about the large-scale structure in 
the area due to the heavy obscuration.  
This seems to be a generally empty region of the sky, judging from optical and 
IR galaxy counts, but it's not clear if this is real.  The 21 cm mapping will 
address this, since it is unaffected by dust and high stellar density.

In the outer Galaxy ZOA, the area
to be probed includes
the Gemini void ($\ell$, b, v = 172$^\circ$, 9$^\circ$, 3000 km s$^{-1}$)
and Monoceros filament, a mass overdensity at low redshift, 
extending into the Arecibo sky at $\ell \sim 210^\circ$ 
(Henning et al.~2005, Kraan-Korteweg et al.~2005).
Also covered will be the Orion void
($\ell$, b, v = 206$^\circ$, -2$^\circ$, 1500 km s$^{-1}$), interesting
because 
of two low velocity galaxies within its putative borders discovered with HIZOA,
suggesting that it
may not be a void at all.
At higher velocities the Gemini void may continue - ALFA ZOA
should determine this clearly.
We will probe part of the Canis Major void ($\ell$, b, v = 229$^\circ$, 
-13$^\circ$, 5000 km s$^{-1}$; Donley \etal (2005) suggest its center may lie at
$\ell$, b = 220$^\circ$,
0$^\circ$.)
The portion of the Pisces-Perseus chain which may connect to 
A569 ($\ell$, b, v = 168$^\circ$, 23$^\circ$, 5900; Fairall 1998),
as surmised by Pantoja et al.~(1997) will be probed where it lies
behind the Milky Way.
In both the inner and outer Galaxy ZOA regions, ALFA ZOA will almost certainly
uncover previously-unrecognized structures, as did its predecessor HIZOA.

In addition to improving our understanding of the local galaxy density field, 
ALFA ZOA will also help to improve our understanding of the local galaxy 
velocity field, and correspondingly, the local {\it matter} density field.  
Because the number of galaxies with known redshifts greatly exceeds the number of 
galaxies with redshift-independent distance indicators, several authors have 
attempted to reconstruct the velocity and matter density fields of the 
local universe using redshifts alone, operating under the assumption that the matter 
density field is related to the galaxy density field by some assumed biasing relation 
(e.g., Branchini \etal 1999, Erdogdu \etal 2006).  
However, because most galaxy surveys do not extend to low Galactic latitudes, 
we are missing an important part of the local velocity/density field.  
Even the near infrared, less affected by dust obscuration than the optical, retains a ZOA.
In reconstructing density and velocity fields from the 2MASS Redshift Survey
(Huchra \etal 2005),
Erdogdu \etal (2006) are forced to deal in a statistical way with missing data in 
the ZOA, defined broadly
to be within $|b|=5^{\circ}$, but flaring significantly toward the Galactic Center.

\section{Precursor Observations}

Before the start of the multi-year survey, we conducted smaller
"precursor" observations, to develop optimal observing and reduction techniques.
Observations were conducted using ALFA\footnote{http://www.naic.edu/alfa/} on the 
305-m radio telescope located at Arecibo, Puerto Rico.  
The ALFA receiver has 7 independent beams, each with two orthogonal linear
polarizations.
The six outer beams are arranged in a hexagonal pattern around the central beam.
At 1.4 GHz, the mean system temperature is 30K, and the mean half-power beam is 3.4 arcmin, 
with the outer beams more 
elliptical in shape than the symmetric central beam (see Fig. 3 of Auld \etal 2006).
For more detailed information on ALFA, see Giovanelli \etal (2005).
Two regions of sky straddling the Galactic plane were observed during these precursor
observations.
A 38-square degree area
near $\ell = 40^\circ$ was observed in June/July 2005 and May 2006.
A 100-square degree region near $\ell = 192^\circ$ was observed in October 2005.
(Fig 1 shows search boundaries).
A third region behind the Taurus molecular cloud was also observed using a similar
mode, though with shorter effective integration time, and during the day.
Results from this area are presented by Lamm \etal (2007), and in a forthcoming paper.

\subsection{Observing Mode}

The observations described in this paper were conducted simultaneously with
GALFA surveys of Galactic HI (Stanimirovic \etal 2006).
While the typical observing mode for extragalactic ALFA surveys is drift scanning,
the observations described here were done in ``nodding'' 
mode.  
In this mode, the receiver either stays on the meridian
and nods up and down in zenith angle, or is positioned at non-zero hour angle, and moves back
and forth in azimuth (the latter technique is used only near declination $18.5^\circ$, near the zenith, 
where nodding on the meridian will not work, due to technical limitations of the telescope).
As the receiver nods and the Earth turns, the telescope traces a zigzag pattern on
the sky as viewed in RA-Dec.
The rotation angle of ALFA is selected to keep the separation between beams
constant during the scans, with
beam spacing of 1.8 arcmin orthogonal to
the scan direction.
Each subsequent day's scanning begins at different LST, such that
the spacing between the edge beams of adjacent scans is the same as that
between beams in the same scan.

Because different days' scans cross each other, each position on the sky is 
observed twice, separated by at least 24 hours.
The repeated observations are used to improve the RFI rejection in the data.
The motion of the receiver and the overlapping observations
lead to an effective integration time of 8 seconds per beam.
Calibration is handled by firing a high-temperature noise diode at the beginning of each 
scan, and at its halfway point.

\subsection{Backend}

While scanning, spectra were recorded every second using the Wide-band
Arecibo Pulsar Processors (WAPPs) as the back-end signal processors.
The WAPPs were configured to cover 100 MHz bandwidth centered at 1383
MHz.  With rolloff in sensitivity at the bandpass extremes, the useful
search range was -1000 \kms to 17,750 \kms.  The 100 MHz band was
divided into 4096 channels, producing a channel spacing of 24 kHz, or
5 \kms in the HI line.  Because of the constant presence of the
strong, narrow Galactic HI signal at zero velocity, which causes
ringing in the spectra, Hanning smoothing was applied in the first
stage of data reduction, increasing the velocity resolution by a
factor of two, to 10 \kms.

\begin{figure*}[ht]
\epsscale{1.5}
\plotone{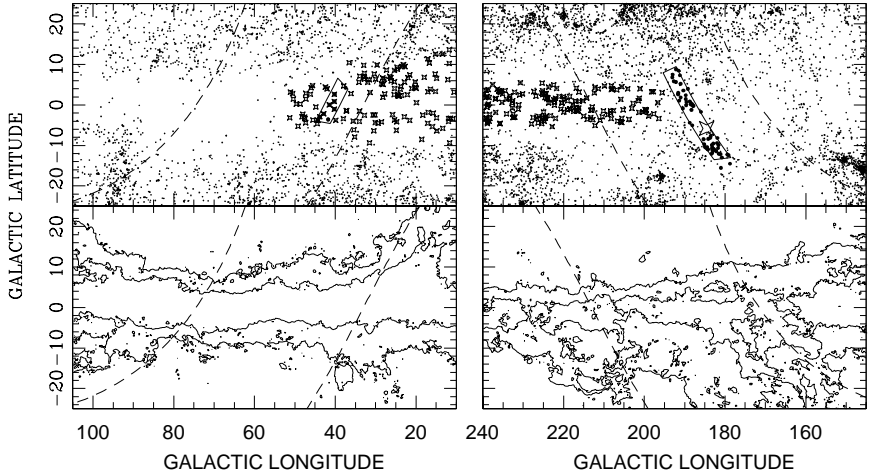}
\caption{Top panels: Sky distribution of cataloged LEDA optical/IR
  objects with velocities within 12000 km/s (small dots) and Parkes
  HIZOA galaxies (crosses; Donley \etal 2005, Shafi 2008, Henning
  \etal 2005, Kraan-Korteweg \etal 2005, and catalog in prep.)
  surrounding the ALFA ZOA precursor regions (roughly rectangular
  boxes).  ALFA ZOA galaxies are shown as large dots.  The star shows
  the location of the Crab nebula.  The dashed lines indicate the
  declination limits for the Arecibo telescope.  Bottom panels:
  Galactic extinction levels of A$_{\rm B}$ = 1 and 3 mags (Schlegel
  \etal 1998) shown as contours.
\label{fig1}}
\end{figure*}

\section{Data Reduction}

The spectral data were bandpass corrected, Hanning smoothed, and
Doppler-corrected using the AIPS++ package LiveData, originally
developed for the Parkes multibeam surveys (Barnes \etal 2001), with
small modifications to accept the Arecibo CIMAFITS format.  For each
beam and polarization, a median bandpass calculated from the scan was
removed.  The relative gains of the ALFA beams (both polarizations)
were taken into account at this stage.  Also at this stage, the system
temperature calibration was applied.  The calibrated spectral data,
weighted according to the beam shape and normalized by the beam
response, were then gridded into cubes with a pixel size of 1 arcmin.
Each output pixel is the median of spectra that lie within 1.5 arcmin
of the center of the pixel.  This step effectively removes outliers,
an efficient method of mitigating transient RFI.  Sources of RFI which
are not transient in time remain in the data, in particular, several
channels surrounding 7900 \kms and 15,600 \kms were ruined by almost
constant RFI (L3 GPS and FAA radar, respectively).  The velocity range
occupied by Galactic HI is within -100 to 100 \kms, with dependence on
Galactic longitude.  High Velocity Clouds are also seen in the data,
with large complexes reaching to $\sim -200$ \kms.  These will be the
subject of a forthcoming paper.  Away from these velocities, and also
away from the small region affected by the Crab Nebula the resulting
rms for the survey was 5.5 - 6 mJy per beam.  Full details of the
bandpass estimation and gridding into datacubes is given by Barnes
\etal (2001).

\subsection{Search Method and Profile Parametrization}

Each three-dimensional (RA-Dec-Vel) datacube was visually inspected
over the entire usable velocity range by three independent searchers,
using the visualization tool Karma KVIS (Gooch 1996).  While
high-latitude HI surveys, such as the HI Parkes All Sky Survey, have
used automatic galaxy detection algorithms to produce galaxy candidate
lists, we have found that the more complicated ZOA produces an
unmanageable number of false detections due to increased continuum
emission at low-Galactic latitude, and that the human eye-brain system
is still far superior for finding galaxies and rejecting spurious
signals.  Lists of galaxy candidates were compiled independently by
each searcher.  Galaxy catalogs in the literature were not consulted
at this stage, to ensure a uniformly HI-selected sample, and not some
complicated function of HI properties and existing observations at
other wavelengths.  When all three searchers agreed, a candidate was
accepted.  In rare cases, when fewer than three searchers had noted a
candidate, positions were re-examined, and sources with at least fives
times the local rms noise over two or more channels were accepted, if
their profile shapes were consistent with known HI objects,
e.g. two-horned, flat-topped, Gaussian, or a combination.

The coordinates and HI parameters of each galaxy were then measured
using the MIRIAD (Sault \etal 1995) task MBSPECT.  MBSPECT fits the
coordinates (centroid) of the HI emission, and then provides a
weighted sum of the emission in each velocity plane to create a
spectral profile.  Visually inspecting each profile to determine a
line-free range, a polynomial (typically a first, second, or third
order polynomial, depending on the baseline shape) is fit to the
baseline region, and then subtracted from the spectrum.  The total
flux is integrated across the line profile in the baseline-subtracted
spectrum, and the widths are computed at 20\% and 50\% of the peak
flux level, using the ``width maximizing'' technique of finding the
outwardmost channels on either side of the profile with fluxes greater
than 20\% and 50\% of the peak flux level respectively.  The systemic
velocity of the galaxy is calculated to be the midpoint of the profile
at the 50\% level.  All of the galaxies are spatially unresolved by
the 3.4 arcmin beam, with the exception of J1901+0651, with 5-6 arcmin
extent.

\subsection{A Posteriori Selection Function}

Because the datacubes were searched visually, there was no hardwired
selection function, such as would be the case with an automatic
galaxy-finding algorithm.  In order to compare the ALFA ZOA selection
function directly to other ALFA blind surveys, namely ALFALFA and
AGES, we show in Figure 2 the detected galaxies' HI flux densities
versus their linewidths.  The dotted line is not a fit to the data
envelope, but rather shows a flux and linewidth dependent
signal-to-noise ratio of 6.5 (following Saintonge (2007) and Cortese
\etal (2008))
$$ S/N = {1000 \times {\rm Flux} \over {W_{50}}} \times {{\rm w^{1/2}} \over {rms}}, $$
where w is either W$_{50}$/$(2 \times \delta v)$ for linewidths less
than 400 \kms, or 400/$(2 \times \delta v)$ for linewidths of 400 \kms
or greater, where $\delta v$ is the velocity resolution of the survey.
(As noted by Saintonge (2007), 400 \kms marks the velocity width at
which typical spectral baseline fluctuations become comparable to the
width of the galaxy profile).  For this survey, the velocity
resolution is 10 \kms, and the rms is taken as the representative
value of 5.75 mJy.  As for ALFALFA and AGES, this linewidth and flux
density dependent signal-to-noise threshold of 6.5 delineates the
selection threshold quite well.

\begin{figure}[th]
\epsscale{1.0}
\plotone{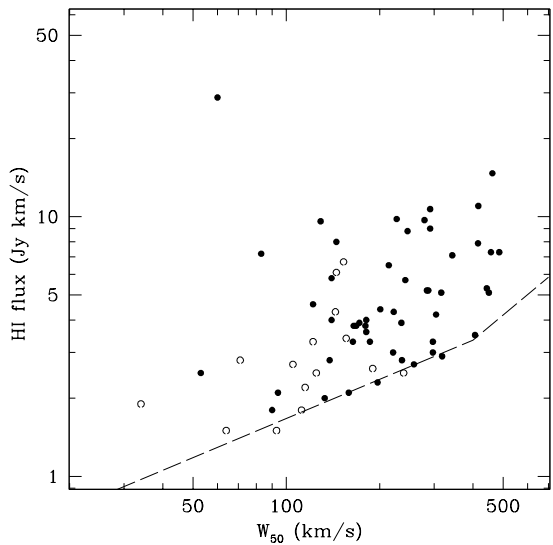}
\caption{Integrated HI flux density versus linewidth at 50\% peak 
level for the ALFA ZOA galaxies. The filled circles represent ALFA ZOA galaxies with counterparts in the literature as described in \S~5.2, the open circles represent ALFA ZOA galaxies with no known counterparts. The dashed line shows a flux density and linewidth dependent reliability limit of S/N = 6.5, as for ALFALFA (Saintonge 2007).  This a posteriori S/N selection threshold for ALFA ZOA is consistent
with the reliability limits of both the ALFALFA and AGES (Cortese \etal 2008) surveys.
\label{fig2}}
\end{figure}

\section{Overview of Galaxies Detected}

\subsection{HI Parameters}

In the two low-latitude precursor regions observed, a total of 72
galaxies were detected: $10$ in the inner Galaxy region (38 square
deg), and $62$ in the outer Galaxy region (100 square deg).  Table 1
presents the HI parameters for these ALFA detections, with columns
containing the following information:

{\it Column (1).}---Source name;

{\it Columns (2) and (3).}---Right ascension and declination (J2000.0)
of the fitted position;

{\it Columns (4) and (5).}---Galactic latitude and longitude of the fitted position;

{\it Column (6).}---The HI flux integral;

{\it Column (7).}---The heliocentric velocity (cz)
taken as the midpoint of the profile at the 50\% level;

{\it Columns (8) and (9).}---The full velocity width of the HI line 
measured at the 50\% and 20\% levels respectively;

{\it Column (10).}---Distance to the galaxy, correcting the velocity to the
Local Group frame: 
$$ v_{LG} = v_{hel} + 300 sin(l) cos(b) $$ and taking H$_0$ = 71 \kms Mpc$^{-1}$;

{\it Column (11).}---Logarithm of the HI mass.

The uncertainties on F$_{\rm HI}$, V$_{\rm hel}$, W$_{50}$, and
W$_{20}$ were calculated following the discussion in Koribalski et al.
(2004).  The error on the HI flux integral is

$$ \sigma(F_{HI}) = 4\;(SN)^{-1}(S_{peak}F_{HI}\delta\nu)^{1/2} \;, $$

where $S_{peak}$ is the peak flux, SN is the signal to noise ratio
$S_{peak}$ to $\sigma(S_{peak})$, $F_{HI}$ is the integrated flux, and
$\delta\nu$ is the velocity resolution of the data, 10 km/s.
$\sigma(S_{peak})$ is the error in the peak flux

$$ \sigma(S_{peak})^2 = rms^2 + (0.05\;S_{peak})^2 \;. $$

The uncertainty in the systemic velocity is
$$ \sigma(V_{hel}) = 3\;(SN)^{-1}(P\delta\nu)^{1/2} \;, $$
with
$$ P = 0.5\;(W_{20}-W_{50}) \;, $$
is a measure of the steepness of the profile edges. 
The uncertainties in the linewidths are given by
$$ \sigma(W_{20}) = 3 \sigma(V_{hel})\;, $$
$$ \sigma(W_{50}) = 2 \sigma(V_{hel})\;. $$

The uncertainties in the HI masses are derived from the uncertainties
in the HI flux integrals.

Included in this table of 72 sources are six galaxies which are
clearly detected, but whose signals lie on the edge of the survey
region.  They are assigned ALFA ZOA names, but since some 21-cm flux
is missing, their positions, flux densities, and heliocentric
velocities are indicative only (and marked with ``:'' in Table 1) and
no further quantities are derived.  Figure 3 presents the HI 21-cm
spectra of the 66 galaxies with secure HI measurements.


\begin{deluxetable}{rrrrrrrrrrr}
\tabletypesize{\scriptsize}
\rotate
\tablecolumns{11}
\tablewidth{0pc}
\tablecaption{\HI Galaxy Parameters from ALFA Zone of Avoidance Precursor Observations}
\tablehead{
\colhead{ALFAZOA} & \colhead{RA} & \colhead{Dec} & \colhead{l} & \colhead{b} & \colhead{F$_{HI}$} & \colhead{V$_{hel}$} 
& \colhead{W$_{50}$} &\colhead{W$_{20}$} & \colhead{D$_{LG}$} & \colhead{log M$_{HI}$} \\

\colhead{} & \colhead{(J2000.0)} & \colhead{(J2000.0)} & \colhead{(deg)} & \colhead{(deg)} & \colhead{(Jy \kms)} & \colhead{(\kms)} &
\colhead{(\kms)} & \colhead{(\kms)} & \colhead{(Mpc)} & \colhead{(\msun)}} 

\startdata
$J1844\!+\!0554$&$18\;44\;34$&$+05\;54\;33$&$37.42$&$4.22$&$1.5\pm0.5$&$9650\pm9$&$64\pm18$&$107\pm27$&$139$&$9.83\pm0.18$\\
$J1855\!+\!0737$&$18\;55\;13$&$+07\;37\;19$&$40.15$&$2.63$&$5.2\pm0.9$&$6153\pm5$&$287\pm10$&$301\pm14$&$89$&$9.99\pm0.08$\\
$J1901\!+\!0651$&$19\;01\;35$&$+06\;51\;42$&$40.19$&$0.88$&$28.8\pm1.6$&$2950\pm3$&$60\pm6$&$96\pm8$&$44$&$10.12\pm0.03$\\
$J1906\!+\!0734$&$19\;06\;41$&$+07\;34\;47$&$41.41$&$0.09$&$3.8\pm0.8$&$3082\pm6$&$165\pm11$&$194\pm17$&$46$&$9.28\pm0.10$\\
$J1908\!+\!0559$&$19\;08\;26$&$+05\;59\;49$&$40.20$&$-1.03$&$3.6\pm0.8$&$4555\pm4$&$181\pm7$&$190\pm11$&$67$&$9.58\pm0.11$\\
$J1917\!+\!0749$&$19\;17\;24$&$+07\;49\;07$&$42.85$&$-2.16$&$6.5\pm0.9$&$3027\pm6$&$214\pm12$&$254\pm19$&$46$&$9.50\pm0.06$\\
$J1920\!+\!0612$&$19\;20\;03$&$+06\;12\;25$&$41.73$&$-3.49$&$1.8\pm0.7$&$6207\pm7$&$112\pm14$&$130\pm22$&$90$&$9.54\pm0.21$\\
$J1920\!+\!0610$&$19\;20\;41$&$+06\;10\;28$&$41.77$&$-3.64$&$7.3\pm1.2$&$6491\pm4$&$486\pm7$&$493\pm11$&$94$&$10.18\pm0.07$\\
$J1922\!+\!0818$&$19\;22\;01$&$+08\;18\;23$&$43.81$&$-2.94$&$5.8\pm0.7$&$3109\pm3$&$140\pm7$&$161\pm10$&$47$&$9.47\pm0.06$\\
$J1925\!+\!0816$&$19\;25\;56$&$+08\;16\;18$&$44.24$&$-3.81$&$1.8\pm0.5$&$3085\pm5$&$90\pm10$&$107\pm16$&$46$&$8.96\pm0.14$\\
$J0449\!+\!2141$&$04\;49\;48$&$+21\;41\;05$&$178.79$&$-14.49$&$8.0\pm0.9$&$3918\pm3$&$145\pm6$&$164\pm9$&$55$&$9.76\pm0.05$\\
$J0451\!+\!1920$&$04\;51\;48$&$+19\;20\;13$&$181.02$&$-15.53$&$4.3\pm0.9$&$4713\pm5$&$144\pm10$&$163\pm15$&$66$&$9.65\pm0.10$\\
$J0457\!+\!2241$&$04\;57\;53$&$+22\;41\;60$&$179.12$&$-12.38$&$3.8\pm0.8$&$5295\pm5$&$168\pm11$&$186\pm16$&$75$&$9.70\pm0.10$\\
$J0459\!+\!2231$&$04\;59\;07$&$+22\;31\;58$&$179.43$&$-12.25$&$2.8\pm0.6$&$5122\pm10$&$71\pm19$&$158\pm29$&$72$&$9.54\pm0.11$\\
$J0459\!+\!2102$&$04\;59\;58$&$+21\;02\;09$&$180.79$&$-12.98$&$4.0\pm0.8$&$5414\pm7$&$140\pm14$&$178\pm20$&$76$&$9.74\pm0.10$\\
$J0503\!+\!2114$&$05\;03\;43$&$+21\;14\;28$&$181.15$&$-12.15$&$2.5\pm0.6$&$1587\pm5$&$125\pm10$&$143\pm16$&$22$&$8.47\pm0.12$\\
$J0505\!+\!2059$&$05\;05\;13$&$+20\;59\;06$&$181.57$&$-12.02$&$3.3\pm0.7$&$5320\pm5$&$164\pm10$&$183\pm15$&$75$&$9.64\pm0.10$\\
$J0508\!+\!2051$&$05\;08\;44$&$+20\;51\;40$&$182.15$&$-11.42$&$2.5\pm0.7$&$6717\pm6$&$239\pm11$&$252\pm17$&$94$&$9.72\pm0.14$\\
$J0509\!+\!1930$&$05\;09\;15$&$+19\;30\;32$&$183.36$&$-12.09$&$4.2\pm0.9$&$7497\pm7$&$304\pm14$&$321\pm21$&$105$&$10.04\pm0.10$\\
$J0509\!+\!1949$&$05\;09\;42$&$+19\;49\;12$&$183.16$&$-11.83$&$4.4\pm0.8$&$5669\pm8$&$201\pm16$&$243\pm23$&$80$&$9.82\pm0.09$\\
$J0510\!+\!2138$&$05\;10\;19$&$+21\;38\;50$&$181.71$&$-10.66$&$5.7\pm0.9$&$4653\pm6$&$242\pm12$&$273\pm17$&$65$&$9.76\pm0.07$\\
$J0510\!+\!2044$&$05\;10\;44$&$+20\;44\;45$&$182.52$&$-11.10$&$5.2\pm1.1$&$7146\pm15$&$284\pm31$&$377\pm46$&$101$&$10.09\pm0.10$\\
$J0511\!+\!2024$&$05\;11\;33$&$+20\;24\;39$&$182.92$&$-11.13$&$3.3\pm0.6$&$3862\pm6$&$122\pm12$&$156\pm17$&$54$&$9.36\pm0.09$\\
$J0511\!+\!2014$&$05\;11\;59$&$+20\;14\;48$&$183.11$&$-11.15$&$3.4\pm0.8$&$5044\pm8$&$156\pm17$&$195\pm25$&$71$&$9.60\pm0.12$\\
$J0512\!+\!2039$&$05\;12\;10$&$+20\;39\;11$&$182.80$&$-10.88$&$4.0\pm0.7$&$3824\pm4$&$181\pm8$&$197\pm12$&$54$&$9.43\pm0.08$\\
$J0512\!+\!2210$&$05\;12\;41$&$+22\;10\;14$&$181.60$&$-9.91$&$2.7\pm0.7$&$9104\pm5$&$258\pm10$&$269\pm15$&$128$&$10.02\pm0.13$\\
$J0513\!+\!2022$&$05\;13\;05$&$+20\;22\;09$&$183.16$&$-10.86$&$2.2\pm0.6$&$7327\pm8$&$115\pm16$&$145\pm24$&$103$&$9.74\pm0.14$\\
$J0513\!+\!2033$&$05\;13\;24$&$+20\;33\;09$&$183.05$&$-10.70$&$2.3\pm0.7$&$5593\pm18$&$197\pm35$&$294\pm53$&$79$&$9.52\pm0.16$\\
$J0514\!+\!2031$&$05\;14\;17$&$+20\;31\;57$&$183.18$&$-10.54$&$7.3\pm1.0$&$8552\pm6$&$457\pm13$&$488\pm19$&$120$&$10.40\pm0.07$\\
$J0515\!+\!1928$&$05\;15\;18$&$+19\;28\;30$&$184.21$&$-10.94$&$2.1\pm0.6$&$5175\pm8$&$159\pm17$&$185\pm25$&$73$&$9.42\pm0.15$\\
$J0515\!+\!2209$&$05\;15\;23$&$+22\;09\;44$&$181.97$&$-9.41$&$4.6\pm0.7$&$8962\pm5$&$122\pm10$&$165\pm16$&$126$&$10.24\pm0.08$\\
$J0515\!+\!1921$&$05\;15\;43$&$+19\;21\;45$&$184.36$&$-10.92$&$7.9\pm1.0$&$5285\pm8$&$415\pm16$&$461\pm24$&$74$&$10.01\pm0.06$\\
$J0516\!+\!2051$&$05\;16\;46$&$+20\;51\;53$&$183.24$&$-9.87$&$3.5\pm0.8$&$6763\pm10$&$406\pm20$&$447\pm31$&$95$&$9.87\pm0.11$\\
$J0517\!+\!2120$&$05\;17\;13$&$+21\;20\;23$&$182.90$&$-9.52$&$2.8\pm0.8$&$6984\pm15$&$236\pm31$&$301\pm46$&$98$&$9.80\pm0.15$\\
$J0517\!+\!1936$&$05\;17\;42$&$+19\;36\;08$&$184.42$&$-10.40$&$11.0\pm1.1$&$5255\pm5$&$416\pm10$&$447\pm15$&$74$&$10.15\pm0.05$\\
$J0518\!+\!1909$&$05\;18\;55\!:$&$+19\;09\;41\!:$&$184.96\!:$&$-10.40\!:$&\nodata$$&$5960\!:$&\nodata$$&\nodata$$&\nodata$$&\nodata$$\\
$J0519\!+\!2256$&$05\;19\;38\!:$&$+22\;56\;39\!:$&$181.87\!:$&$-8.15\!:$&\nodata$$&$7180\!:$&\nodata$$&\nodata$$&\nodata$$&\nodata$$\\
$J0521\!+\!1923$&$05\;21\;45$&$+19\;23\;59$&$185.13$&$-9.72$&$3.3\pm0.6$&$4361\pm6$&$186\pm13$&$222\pm19$&$61$&$9.46\pm0.09$\\
$J0524\!+\!2129$&$05\;24\;00$&$+21\;29\;03$&$183.67$&$-8.13$&$4.3\pm0.8$&$5681\pm9$&$222\pm19$&$281\pm28$&$80$&$9.81\pm0.09$\\
$J0524\!+\!2055$&$05\;24\;17$&$+20\;55\;30$&$184.17$&$-8.38$&$1.9\pm0.4$&$7522\pm4$&$34\pm7$&$59\pm11$&$106$&$9.70\pm0.10$\\
$J0525\!+\!2151$&$05\;25\;28$&$+21\;51\;22$&$183.54$&$-7.64$&$10.7\pm1.1$&$5624\pm5$&$291\pm10$&$336\pm15$&$79$&$10.20\pm0.05$\\
$J0526\!+\!1957$&$05\;26\;34$&$+19\;57\;55$&$185.28$&$-8.46$&$3.9\pm0.8$&$5630\pm4$&$235\pm8$&$245\pm12$&$79$&$9.76\pm0.10$\\
$J0545\!+\!1925$&$05\;45\;12$&$+19\;25\;13$&$188.09$&$-5.03$&$9.7\pm1.1$&$8319\pm6$&$279\pm11$&$322\pm17$&$117$&$10.49\pm0.05$\\
$J0548\!+\!1911$&$05\;48\;55\!:$&$+19\;11\;58\!:$&$188.73\!:$&$-4.39\!:$&\nodata$$&$5780\!:$&\nodata$$&\nodata$$&\nodata$$&\nodata$$\\
$J0552\!+\!2255$&$05\;52\;47\!:$&$+22\;55\;47\!:$&$185.98\!:$&$-1.71\!:$&\nodata$$&$910\!:$&\nodata$$&\nodata$$&\nodata$$&\nodata$$\\
$J0559\!+\!2109$&$05\;59\;45$&$+21\;09\;10$&$188.33$&$-1.21$&$5.1\pm1.1$&$8662\pm10$&$450\pm20$&$488\pm30$&$121$&$10.25\pm0.11$\\
$J0602\!+\!2204$&$06\;02\;13$&$+22\;04\;10$&$187.81$&$-0.25$&$2.0\pm0.6$&$2461\pm7$&$133\pm14$&$159\pm22$&$34$&$8.74\pm0.15$\\
$J0602\!+\!2201$&$06\;02\;36$&$+22\;01\;38$&$187.89$&$-0.20$&$2.8\pm0.7$&$2599\pm5$&$138\pm10$&$152\pm15$&$36$&$8.93\pm0.13$\\
$J0604\!+\!2201$&$06\;04\;55$&$+22\;01\;24$&$188.16$&$0.27$&$5.1\pm0.8$&$8811\pm5$&$316\pm11$&$348\pm16$&$124$&$10.26\pm0.07$\\
$J0605\!+\!2141$&$06\;05\;06$&$+21\;41\;12$&$188.48$&$0.14$&$2.6\pm0.7$&$8933\pm8$&$190\pm15$&$218\pm23$&$125$&$9.98\pm0.14$\\
$J0605\!+\!1927$&$06\;05\;28$&$+19\;27\;18$&$190.46$&$-0.88$&$9.0\pm1.1$&$5776\pm9$&$291\pm17$&$372\pm26$&$81$&$10.14\pm0.06$\\
$J0606\!+\!2036$&$06\;06\;46$&$+20\;36\;23$&$189.61$&$-0.05$&$3.0\pm0.8$&$8801\pm7$&$297\pm15$&$314\pm22$&$123$&$10.03\pm0.13$\\
$J0607\!+\!2152$&$06\;07\;02$&$+21\;52\;23$&$188.53$&$0.62$&$7.1\pm1.0$&$5825\pm6$&$343\pm12$&$375\pm18$&$81$&$10.04\pm0.06$\\
$J0608\!+\!2149$&$06\;08\;28$&$+21\;49\;36$&$188.73$&$0.89$&$6.1\pm0.8$&$2463\pm3$&$145\pm6$&$161\pm9$&$34$&$9.22\pm0.06$\\
$J0609\!+\!2101$&$06\;09\;39$&$+21\;01\;08$&$189.58$&$0.74$&$6.7\pm0.8$&$2547\pm3$&$153\pm6$&$170\pm9$&$35$&$9.29\pm0.06$\\
$J0612\!+\!2113$&$06\;12\;44$&$+21\;13\;14$&$189.74$&$1.47$&$3.9\pm0.8$&$11726\pm13$&$172\pm27$&$291\pm40$&$164$&$10.40\pm0.10$\\
$J0614\!+\!1934$&$06\;14\;31$&$+19\;34\;57$&$191.38$&$1.05$&$3.0\pm0.8$&$13506\pm8$&$221\pm17$&$253\pm25$&$189$&$10.40\pm0.13$\\
$J0616\!+\!2142$&$06\;16\;12$&$+21\;42\;27$&$189.70$&$2.41$&$5.3\pm1.0$&$11471\pm13$&$443\pm26$&$499\pm38$&$161$&$10.51\pm0.09$\\
$J0616\!+\!2019$&$06\;16\;43$&$+20\;19\;06$&$190.98$&$1.85$&$2.7\pm0.6$&$4441\pm6$&$105\pm11$&$133\pm17$&$62$&$9.39\pm0.11$\\
$J0620\!+\!2008$&$06\;20\;53$&$+20\;08\;29$&$191.60$&$2.63$&$9.6\pm0.9$&$1320\pm2$&$129\pm4$&$143\pm6$&$18$&$8.85\pm0.04$\\
$J0621\!+\!2010$&$06\;21\;06$&$+20\;10\;16$&$191.60$&$2.69$&$2.1\pm0.6$&$2272\pm6$&$94\pm11$&$113\pm17$&$31$&$8.68\pm0.15$\\
$J0621\!+\!2256$&$06\;21\;09\!:$&$+22\;56\;00\!:$&$189.16\!:$&$3.99\!:$&\nodata$$&$4590\!:$&\nodata$$&\nodata$$&\nodata$$&\nodata$$\\
$J0621\!+\!2142$&$06\;21\;49$&$+21\;42\;56$&$190.31$&$3.56$&$3.8\pm0.8$&$4743\pm6$&$180\pm13$&$207\pm19$&$66$&$9.59\pm0.10$\\
$J0622\!+\!2019$&$06\;22\;07$&$+20\;19\;49$&$191.57$&$2.98$&$1.5\pm0.6$&$2246\pm9$&$93\pm18$&$114\pm27$&$31$&$8.53\pm0.22$\\
$J0629\!+\!2151$&$06\;29\;37$&$+21\;51\;33$&$191.02$&$5.24$&$2.5\pm0.5$&$1600\pm4$&$53\pm8$&$78\pm12$&$22$&$8.44\pm0.10$\\
$J0633\!+\!2102$&$06\;33\;35$&$+21\;02\;13$&$192.18$&$5.68$&$14.7\pm1.3$&$5459\pm4$&$462\pm8$&$490\pm12$&$76$&$10.30\pm0.04$\\
$J0635\!+\!2039$&$06\;35\;13$&$+20\;39\;55$&$192.68$&$5.86$&$8.8\pm1.0$&$4337\pm2$&$246\pm5$&$257\pm7$&$60$&$9.88\pm0.05$\\
$J0638\!+\!2238$&$06\;38\;03$&$+22\;38\;58$&$191.19$&$7.33$&$7.2\pm0.7$&$1370\pm4$&$83\pm8$&$144\pm11$&$18$&$8.76\pm0.04$\\
$J0639\!+\!2044$&$06\;39\;38$&$+20\;44\;33$&$193.08$&$6.81$&$2.9\pm0.8$&$5370\pm6$&$318\pm11$&$328\pm17$&$75$&$9.58\pm0.14$\\
$J0642\!+\!2108$&$06\;42\;03$&$+21\;08\;18$&$192.98$&$7.50$&$3.3\pm0.8$&$5379\pm5$&$297\pm10$&$306\pm16$&$75$&$9.64\pm0.12$\\
$J0643\!+\!2252$&$06\;43\;10\!:$&$+22\;52\;52\!:$&$191.51\!:$&$8.49\!:$&\nodata$$&$1285\!:$&\nodata$$&\nodata$$&\nodata$$&\nodata$$\\
$J0645\!+\!2225$&$06\;45\;42$&$+22\;25\;38$&$192.18$&$8.82$&$9.8\pm1.0$&$4471\pm5$&$227\pm10$&$275\pm14$&$62$&$9.95\pm0.05$\\
\enddata
\end{deluxetable}

\begin{figure*}[h]
\epsscale{1.8}
\plotone{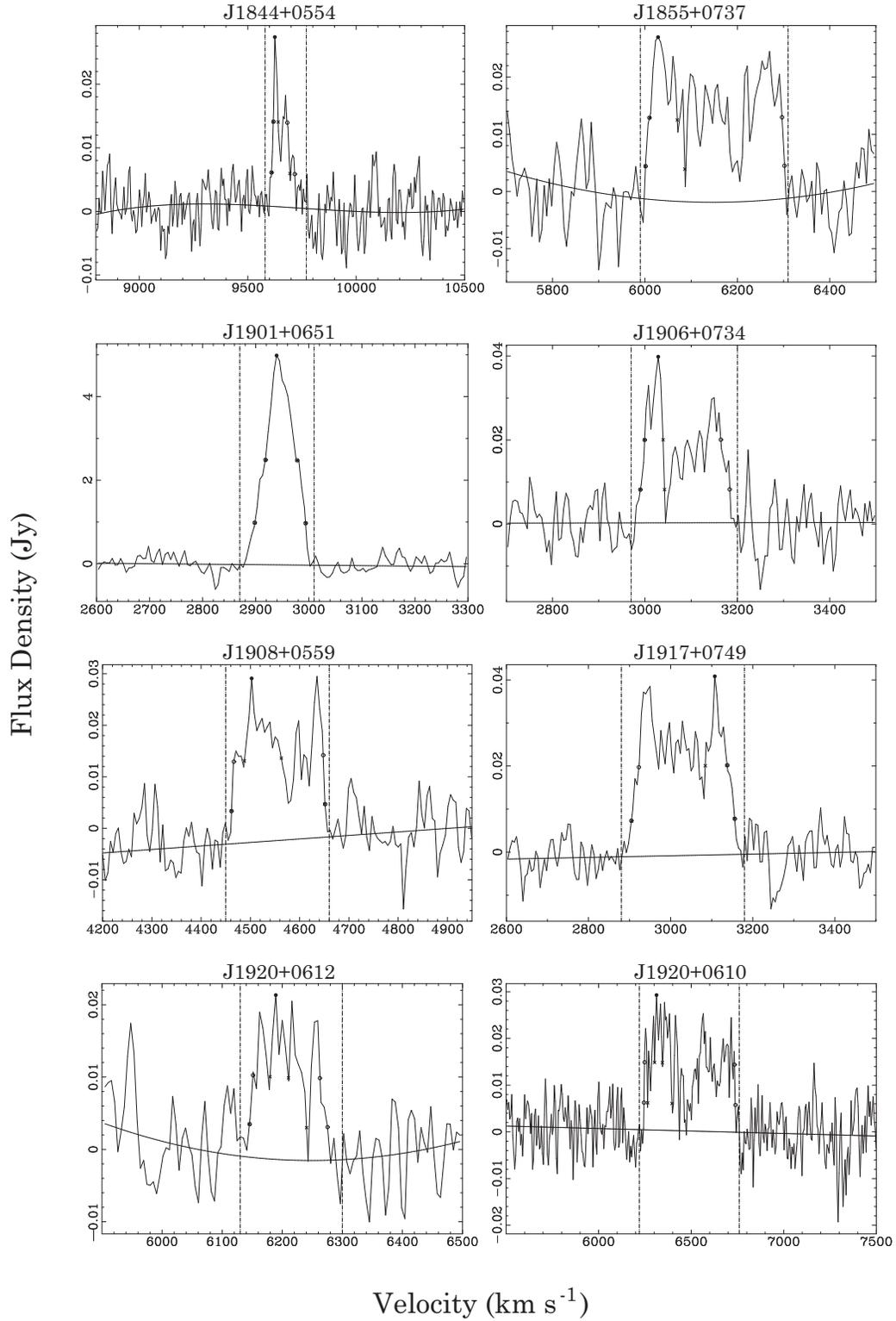}
\caption{ HI spectra of the ALFA ZOA galaxies.  The dotted lines enclose the range in velocity
over which the profile was measured.  The solid line indicates the baseline subtracted.
The small circles on each profile indicate, from top to bottom, the peak, the 50\%, and the 20\% of
peak flux levels.
\label{fig3}}
\end{figure*}

\addtocounter{figure}{-1}

\begin{figure*}
\epsscale{1.8}
\plotone{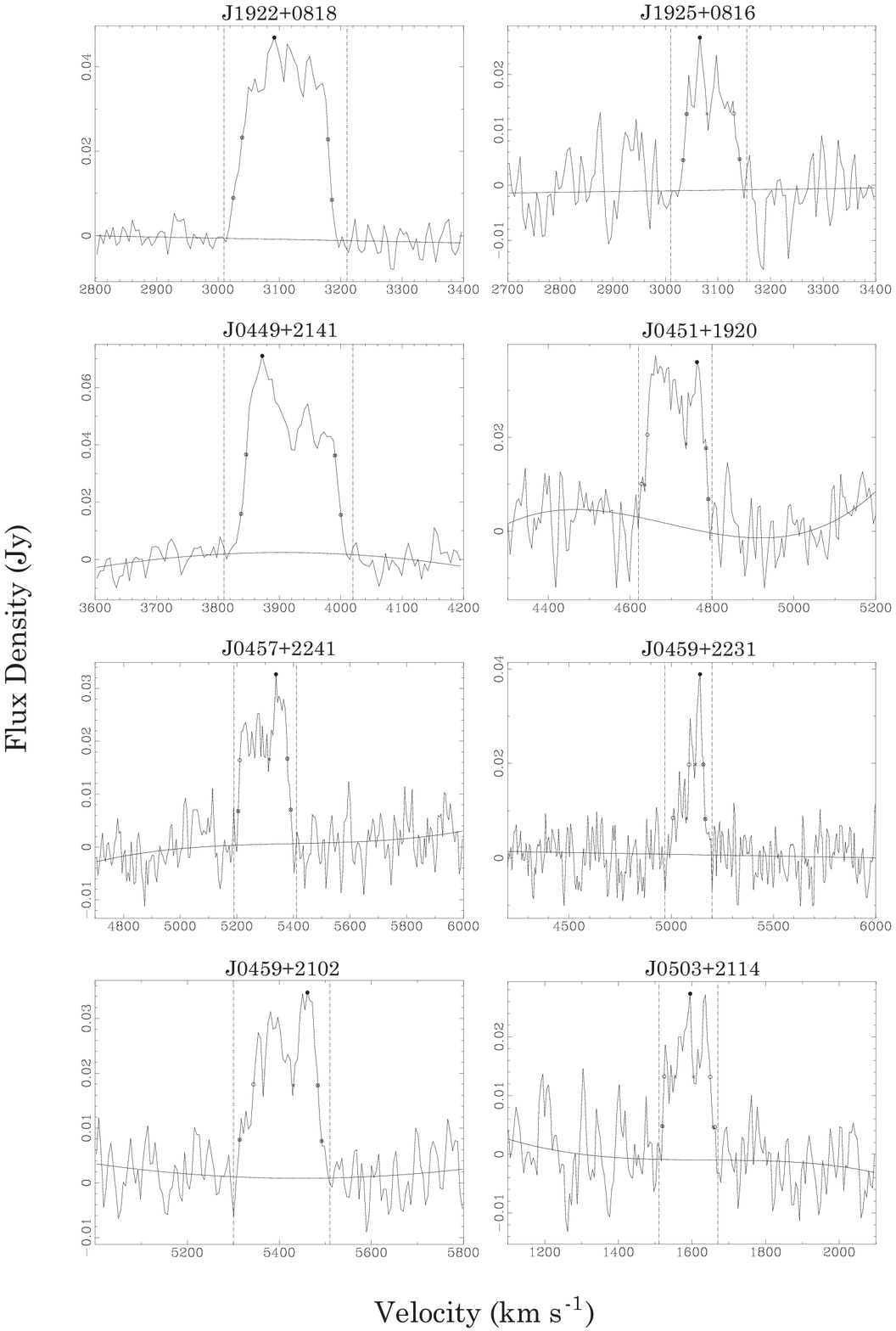}
\caption{(cont.)
\label{fig3}}
\end{figure*}

\addtocounter{figure}{-1}

\begin{figure*}
\epsscale{1.8}
\plotone{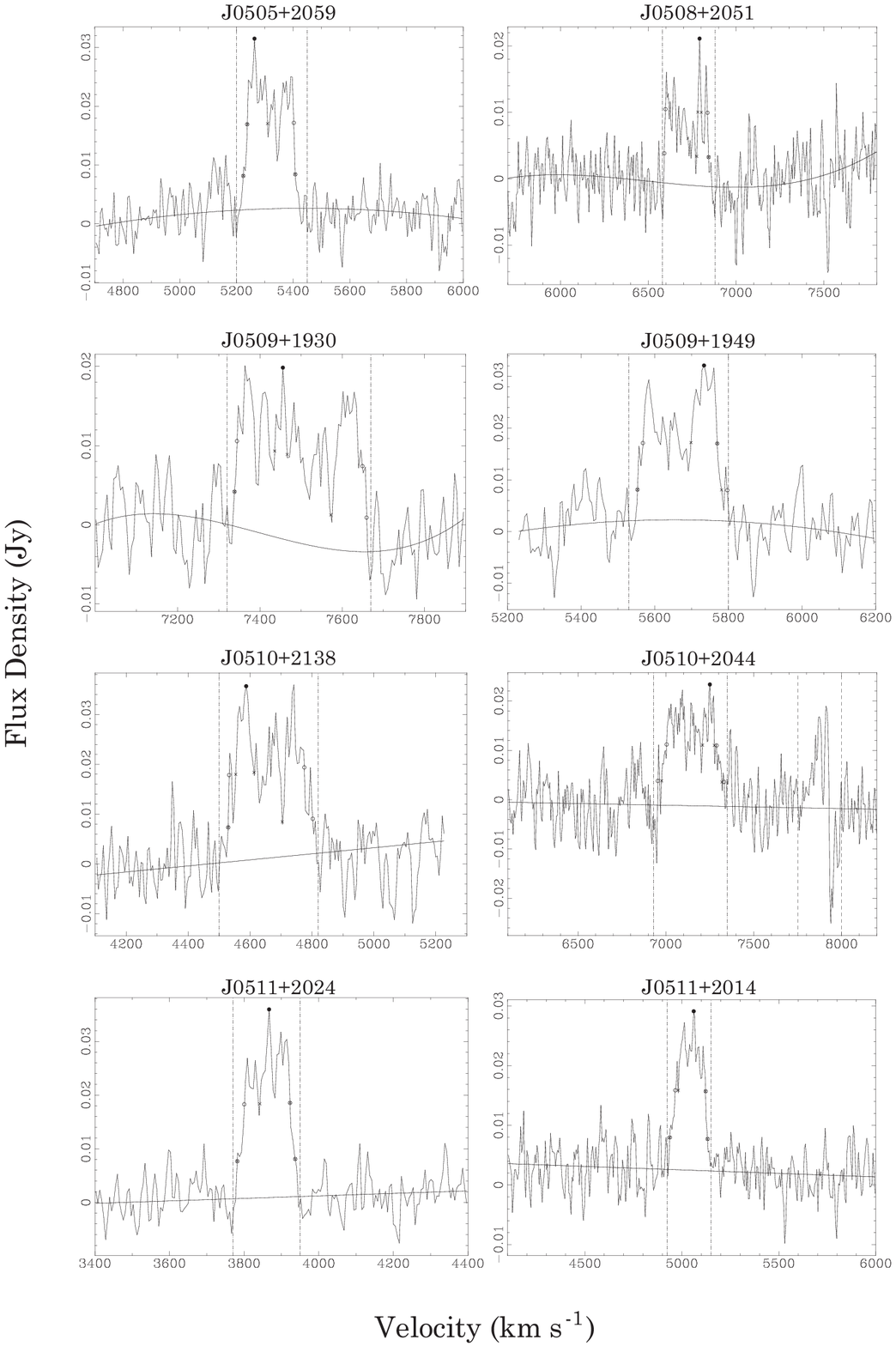}
\caption{(cont.) The portion of the spectrum near 7900 \kms in the panel
for J0510+2044 is affected by RFI and is not included in the baseline fit.
\label{fig3}}
\end{figure*}

\addtocounter{figure}{-1}

\begin{figure*}
\epsscale{1.8}
\plotone{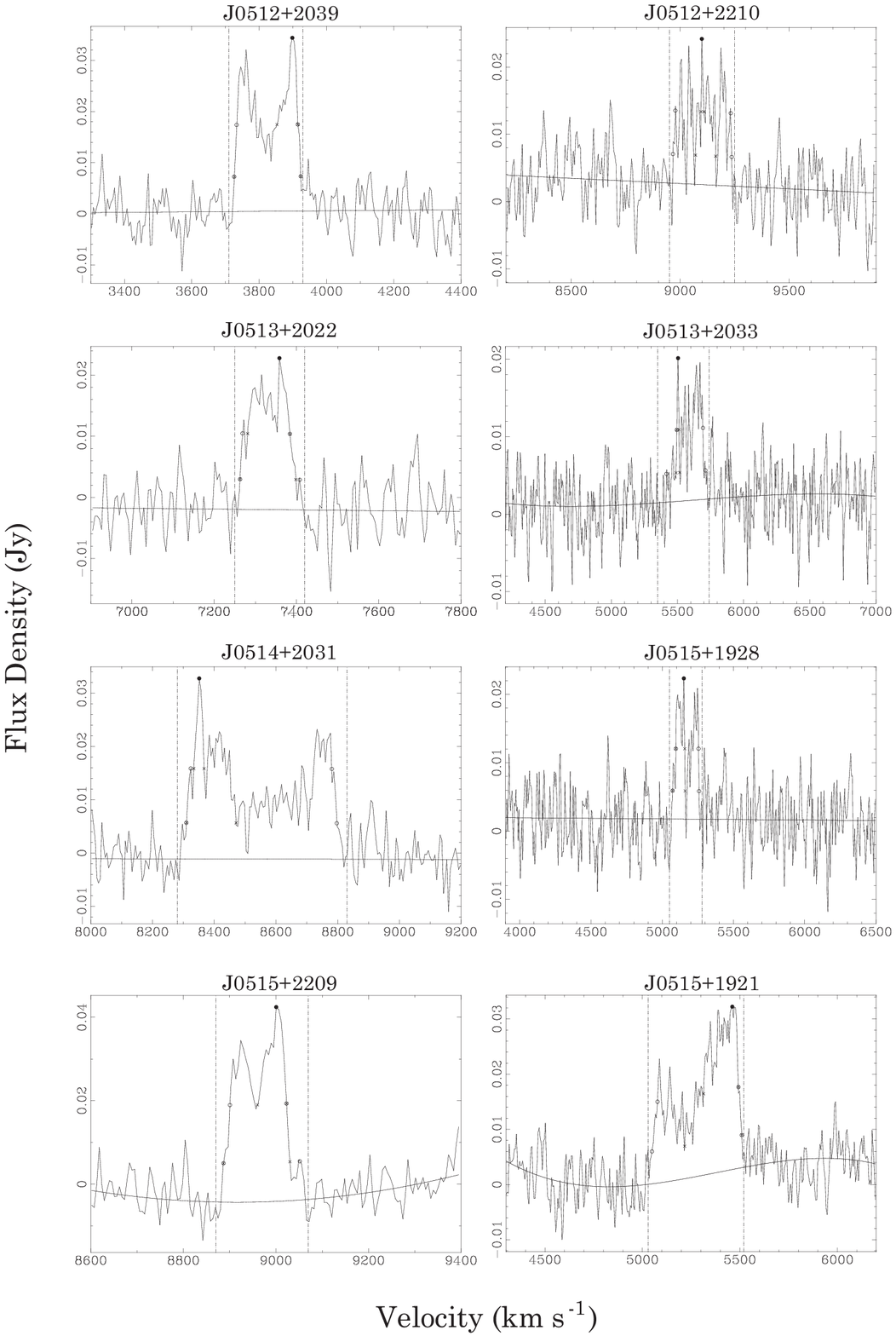}
\caption{(cont.)
\label{fig3}}
\end{figure*}

\addtocounter{figure}{-1}

\begin{figure*}
\epsscale{1.8}
\plotone{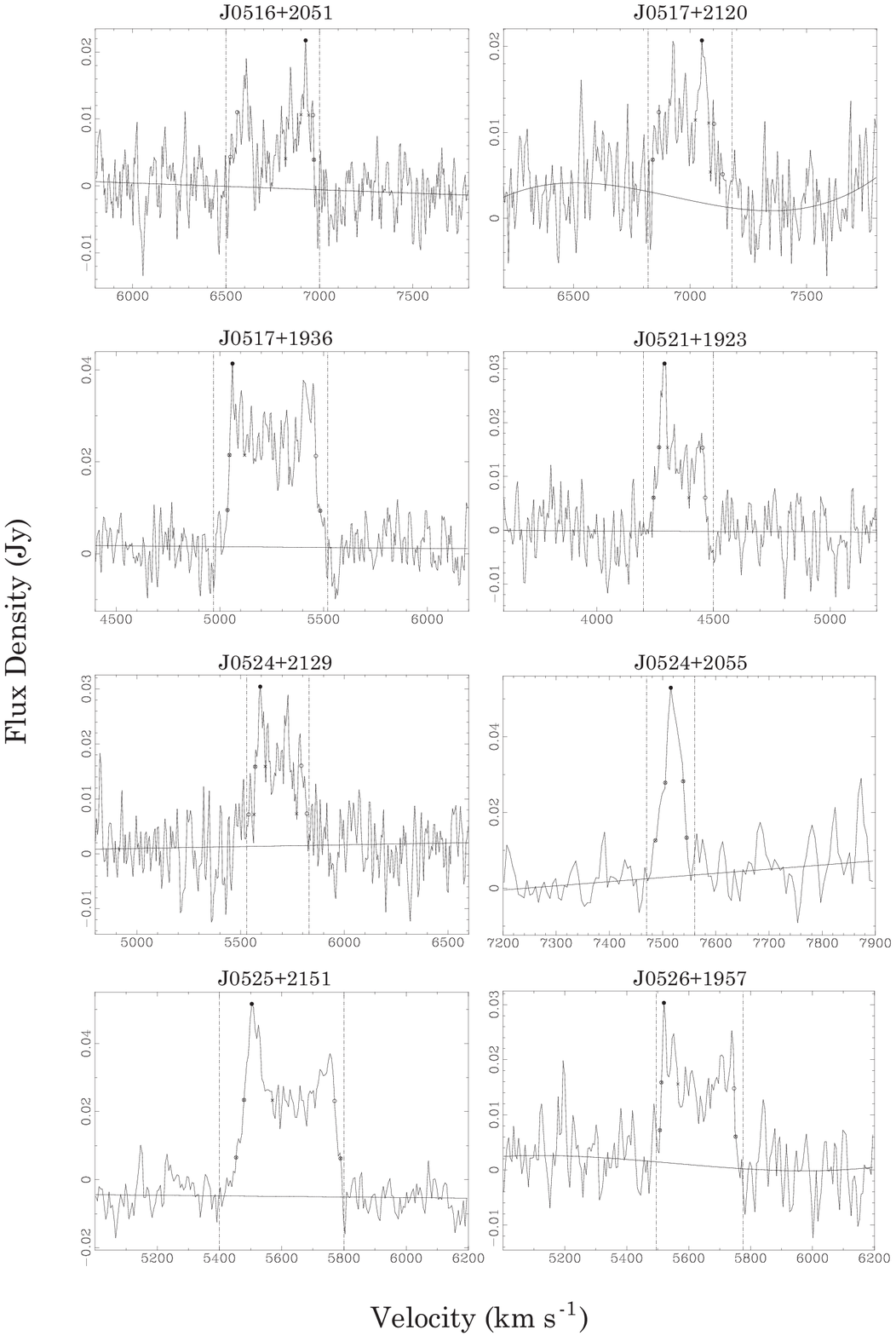}
\caption{(cont.)
\label{fig3}}
\end{figure*}

\addtocounter{figure}{-1}

\begin{figure*}
\epsscale{1.8}
\plotone{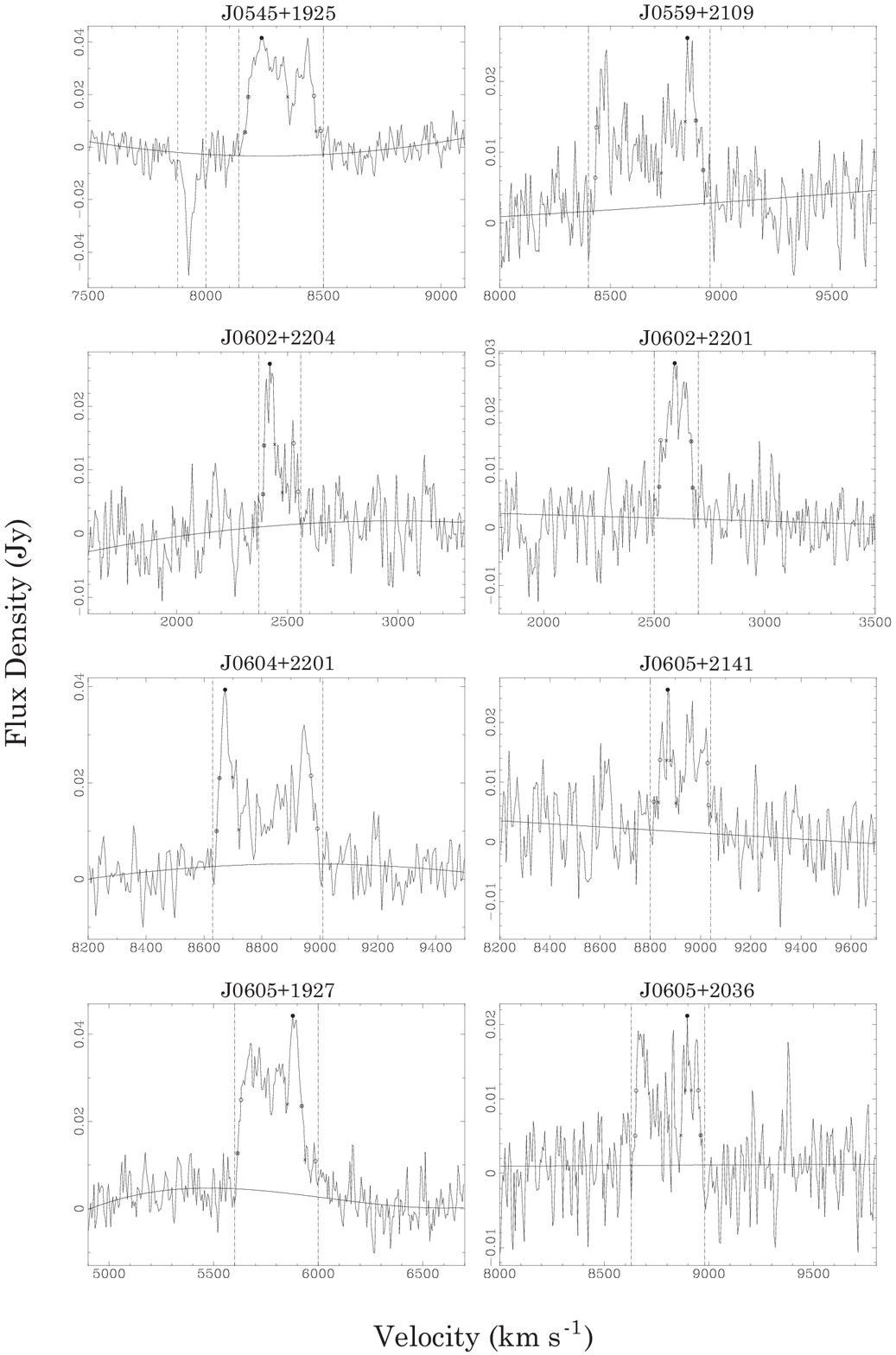}
\caption{(cont.)  The apparent absorption feature near 7900 \kms in the panel for J0545+1925 is
due to RFI, and is excluded from the baseline fit.
\label{fig3}}
\end{figure*}

\addtocounter{figure}{-1}

\begin{figure*}
\epsscale{1.8}
\plotone{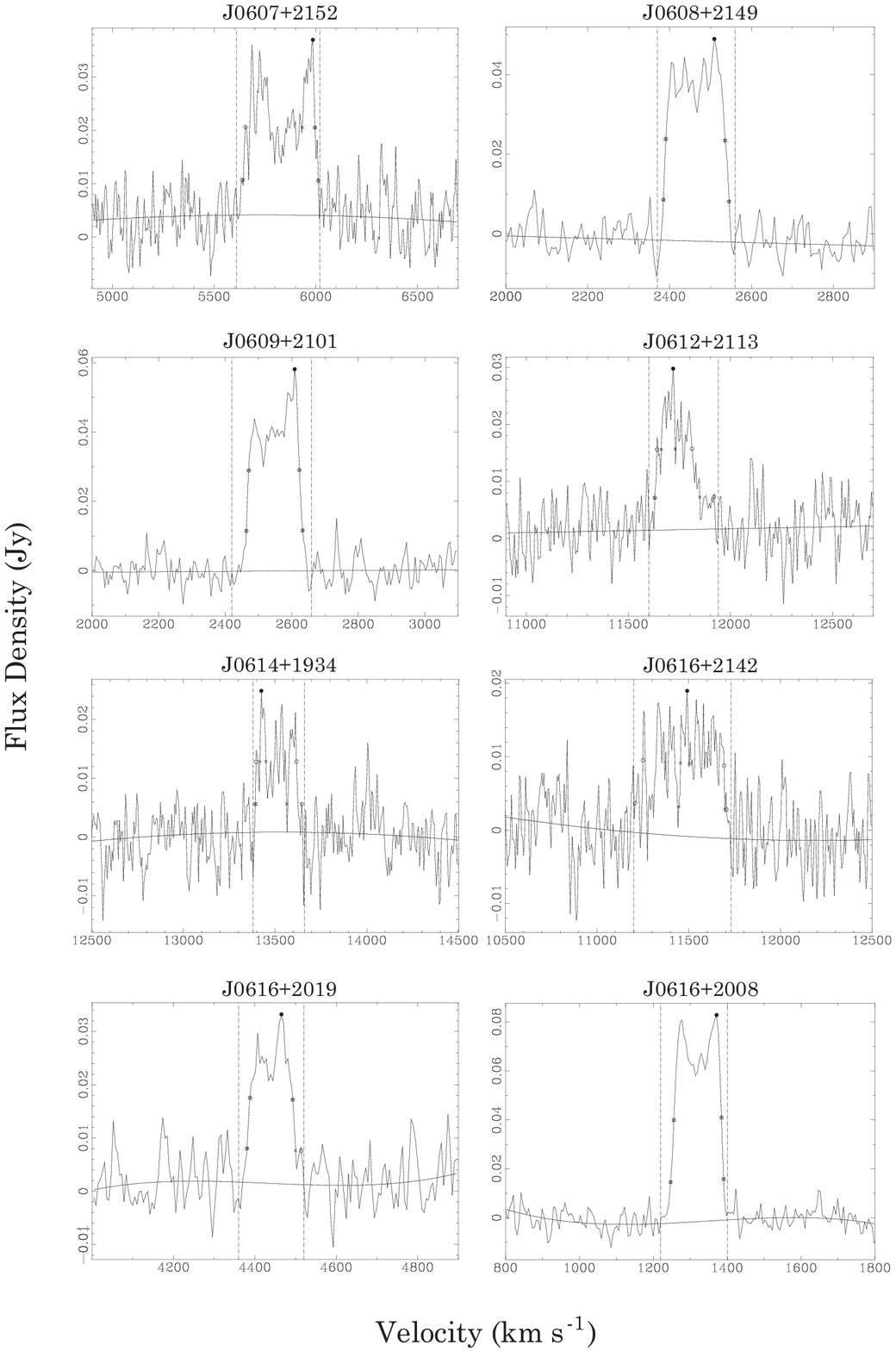}
\caption{(cont.)
\label{fig3}}
\end{figure*}

\addtocounter{figure}{-1}

\begin{figure*}
\epsscale{1.8}
\plotone{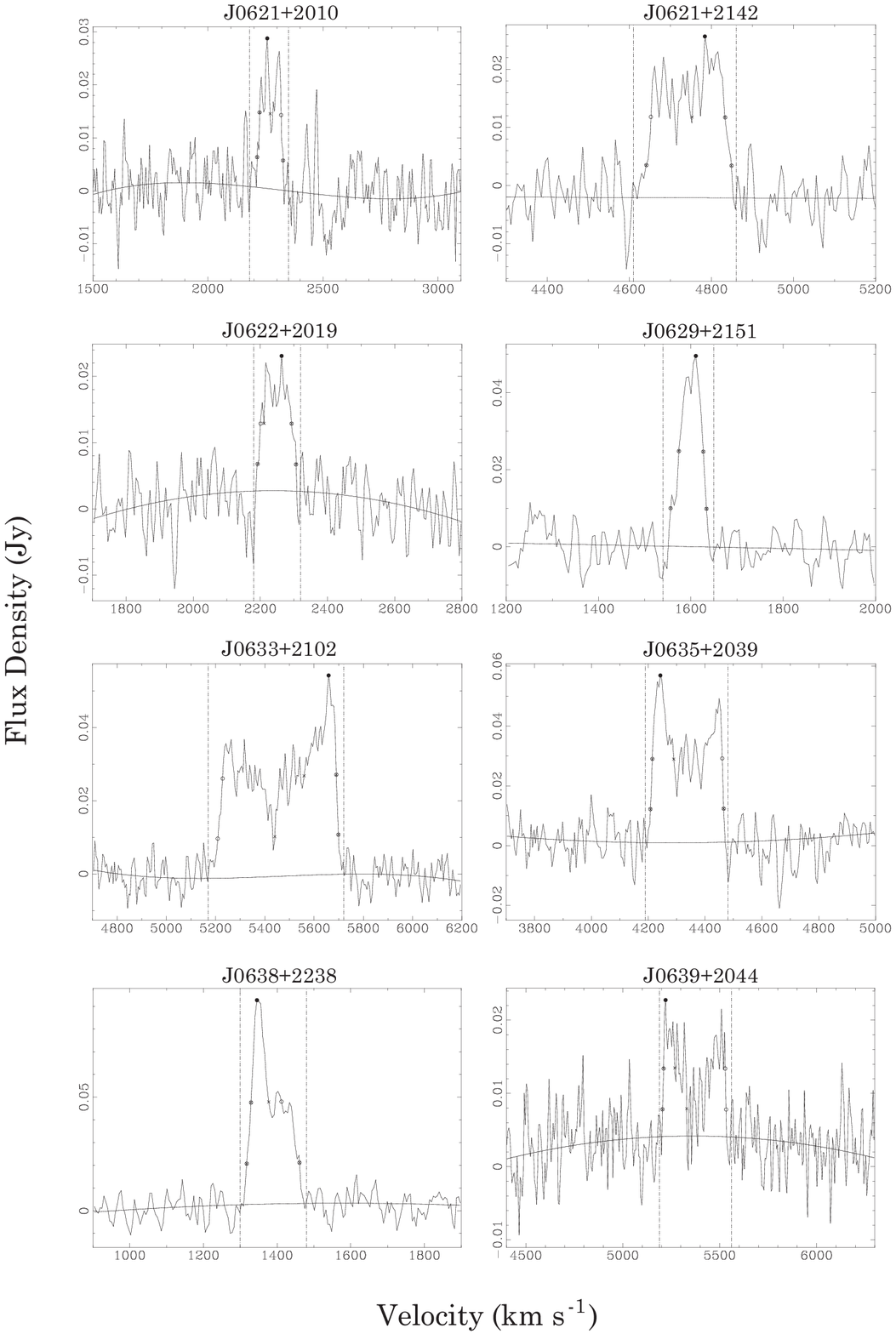}
\caption{(cont.)
\label{fig3}}
\end{figure*}

\addtocounter{figure}{-1}

\begin{figure*}
\epsscale{1.8}
\plotone{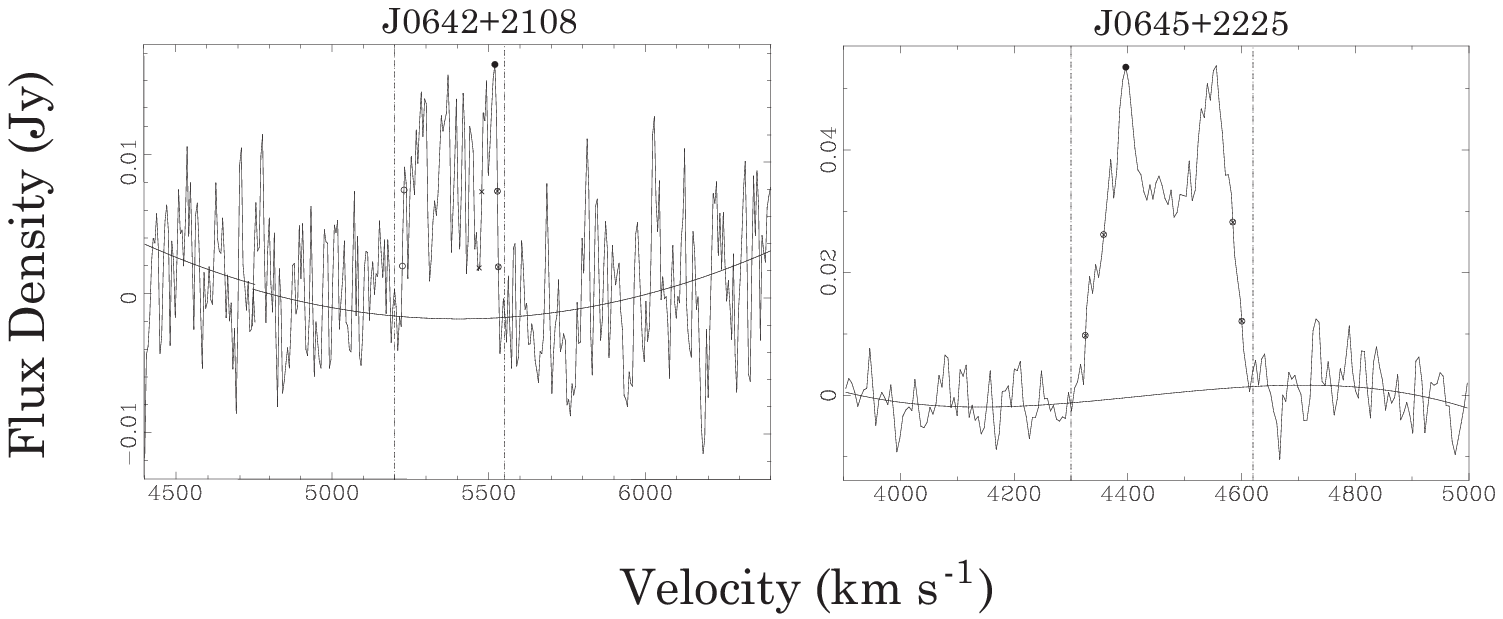}
\caption{(cont.)
\label{fig3}}
\end{figure*}
\clearpage

\begin{figure}[ht]
\epsscale{1.0}
\plotone{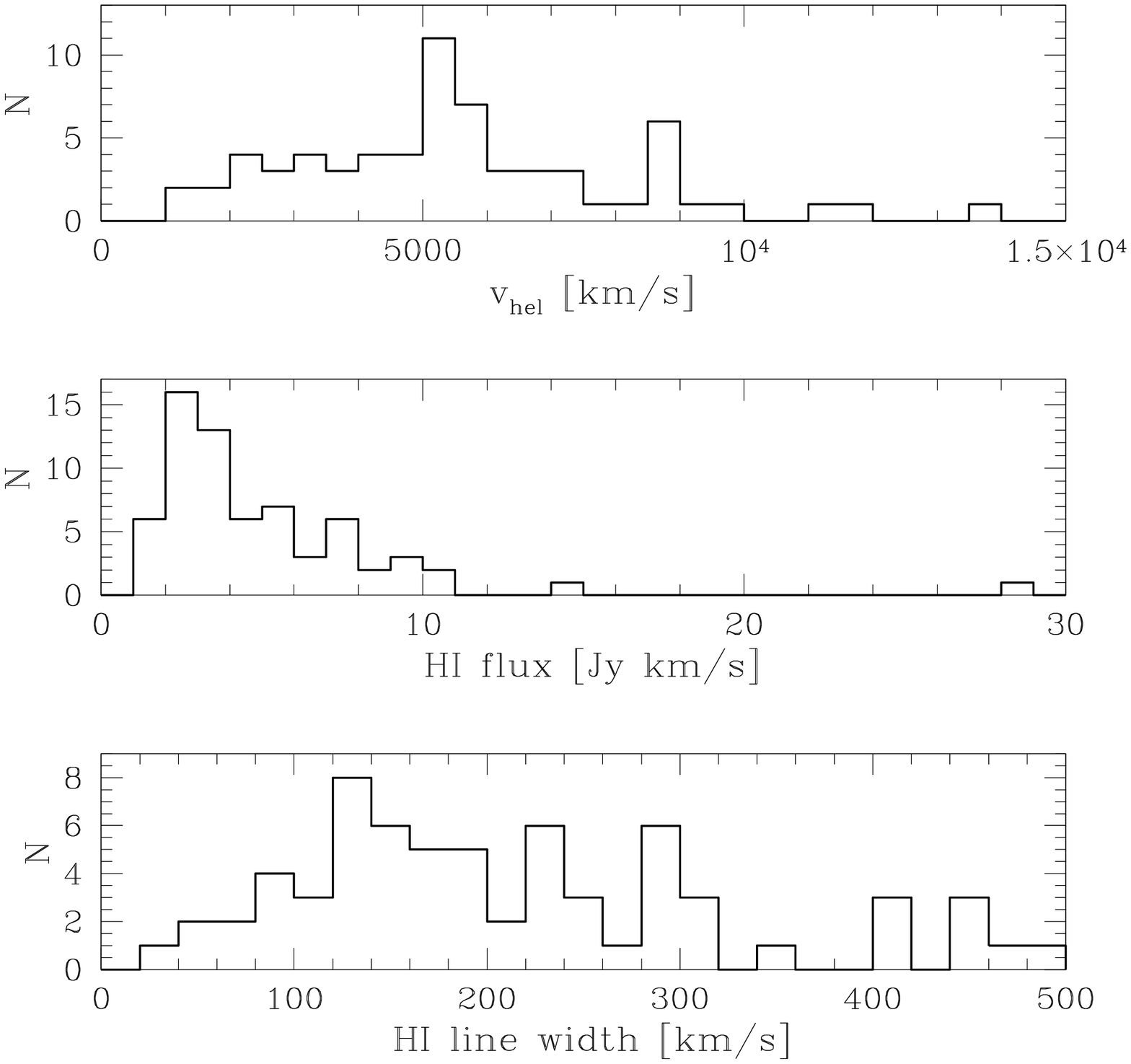}
\caption{Distribution of basic HI parameters for the 66 ALFA ZOA galaxies
with secure measurements.  Top:  systemic velocity.  Middle:  integrated flux
density.  Bottom:  linewidth at 50\% peak.\label{fig4}}
\end{figure}

Figure 4 shows histograms of heliocentric velocity (top panel), HI
flux density (middle panel), and HI linewidths at 50\% of peak flux
(bottom panel) for these objects.  The distribution in heliocentric
velocity, while averaged over some unrelated large-scale structures,
shows an overdensity between 5000 - 6000 \kms due to galaxy
distribution in the outer Galaxy region (described in Section 5.4).
The HI flux density distribution shows that the survey becomes
insensitive below about 2 Jy \kms.  The average linewidth $W_{50}$ for
the survey is 214 \kms with considerable spread, ranging from the
narrowest source with 34 \kms linewidth, to the widest profile, with
486 \kms.

The values of HI mass range from $2.7 \times 10^8$ \msun to $3.2
\times 10^{10}$ \msun, with a mean HI mass of $8.0 \times 10^9$ \msun.
Figure 5 shows the mass distribution of these 66 galaxies with secure
HI measurements.  The mean value and distribution in HI mass are
consistent with expectation, eg. they are very similar to those of the
HIZOA, which was conducted at similar sensitivity and depth.

\begin{figure}[ht]
\epsscale{.80}
\plotone{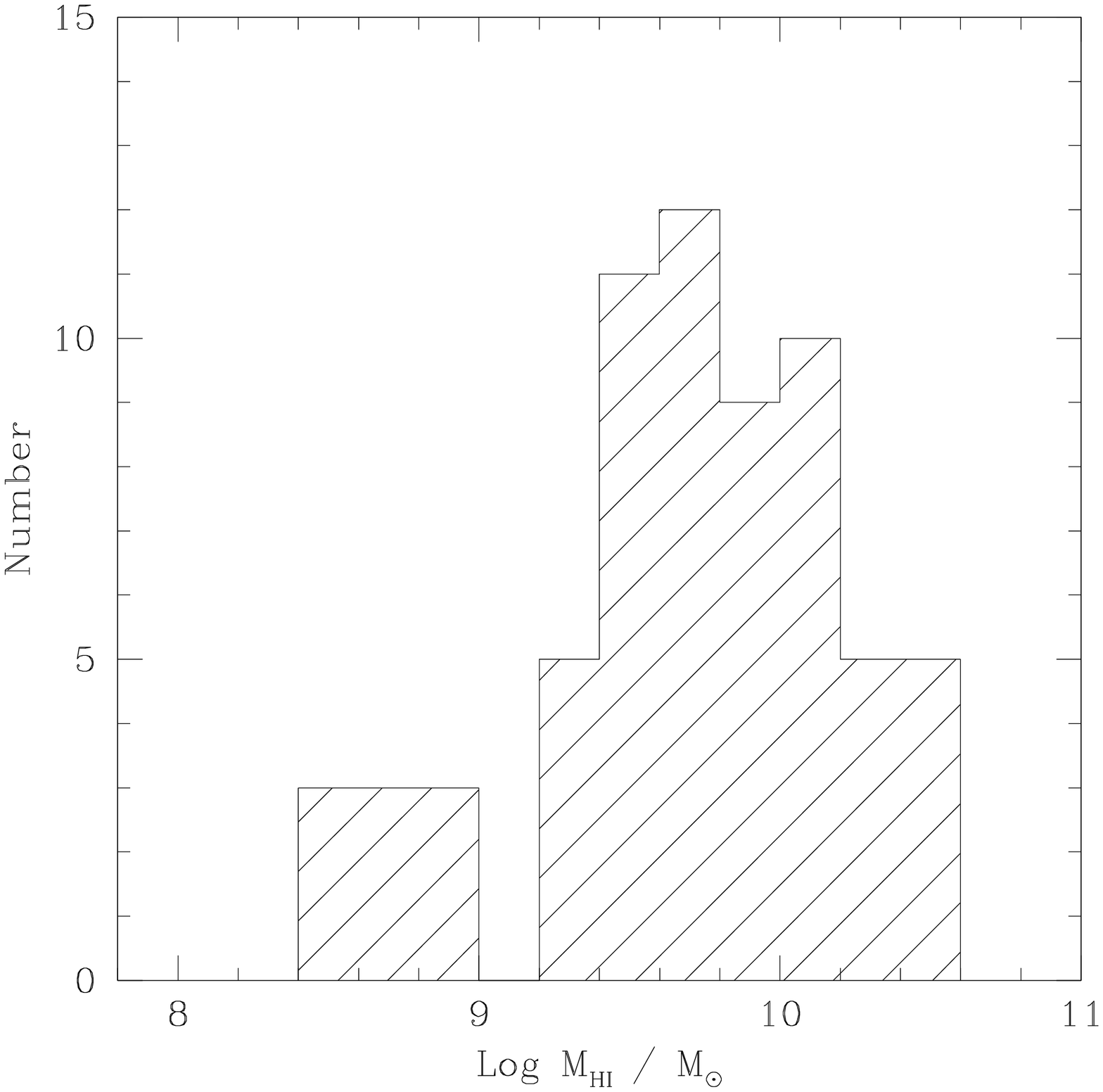}
\caption{ Distribution of HI mass for the survey galaxies.\label{fig5}}
\epsscale{1.0}
\end{figure}

\subsection{Counterparts at Other Wavelengths}

For each ALFA ZOA detection, the NASA Extragalactic Database (NED) was
searched for cataloged galaxies at the location of the HI detection.
The criterion for listing as a possible counterpart is a position
within 3.5 arcmin of the fitted HI position.  No velocity measurement
of the counterpart is necessary for inclusion, but if a velocity is
given, it must be within 100 \kms, otherwise it is excluded as a
possible counterpart.  Also, comparison was made to a recent pointed
HI survey of 2MASS galaxies, which had one object in common with our
survey (van Driel \etal 2009).  Table 2 presents the 57 ALFA ZOA
galaxies with possible counterparts as defined in this way, and
includes:

{\it Column (1).}---Source name (an asterisk indicates the galaxy lies on an edge of the survey
region, so parameters are uncertain);

{\it Column (2).}---Galactic latitude;

{\it Column (3).}---Foreground extinction A$_{\rm B}$ estimated by the
IRAS DIRBE maps of Schlegel \etal (1998);

{\it Column (4).}---Name of HI literature counterpart, if any;

{\it Columns (5) and (6). }---Angular separation,  and velocity difference from HI counterpart
(Literature - ALFA ZOA);

{\it Columns (7) and (8).}---Optical counterpart, if any, and angular separation;

{\it Columns (9) and (10).}---2MASS counterpart, if any, and angular separation;

{\it Columns (11) and (12).}---IRAS counterpart, if any, and angular separation;

{\it Column (13).}--Velocity difference between the ALFA ZOA source and any non-HI velocity measurement
(Literature - ALFA ZOA).

Some galaxies have more than one possible counterpart; no attempt is
made to judge amongst the candidates.  For the galaxies which have
21-cm measurements in the literature [from NED, and Donley \etal
(2005), Lu \etal (1990), Pantoja \etal (1997), Rosenberg and Schneider
(2000), van Driel \etal (2009), Wong \etal (2006)], comparison between
ALFA ZOA and literature values for the heliocentric systemic
velocities, HI flux densities, and linewidths show good agreement
(Fig. 6).

To assess the completeness of the survey, NED was queried to check for
galaxies in these regions which have 21-cm redshifts in the
literature, but were not detected by ALFA ZOA.  In the $\sim140$
square degrees, there are only three galaxies with literature 21-cm
measurements that would indicate they are above our detection limit,
but which were not found in the ALFA ZOA cubes.  One of these objects
appears in the inner Galaxy region, HIZOA J1843+06 (Donley \etal
2005).  Two are in the outer Galaxy region, ADBS J053017+2233
(Rosenberg \& Schneider 2000), and IRAS 05223+1908 (Lu \etal 1990).
We re-examined the ALFA ZOA data at the locations of these sources,
and while they should have been clear detections according to
published values, we do not recover them in our data.  From the
available information, we cannot determine the reason for these three
non-detections.

\begin{figure}[ht]
\plotone{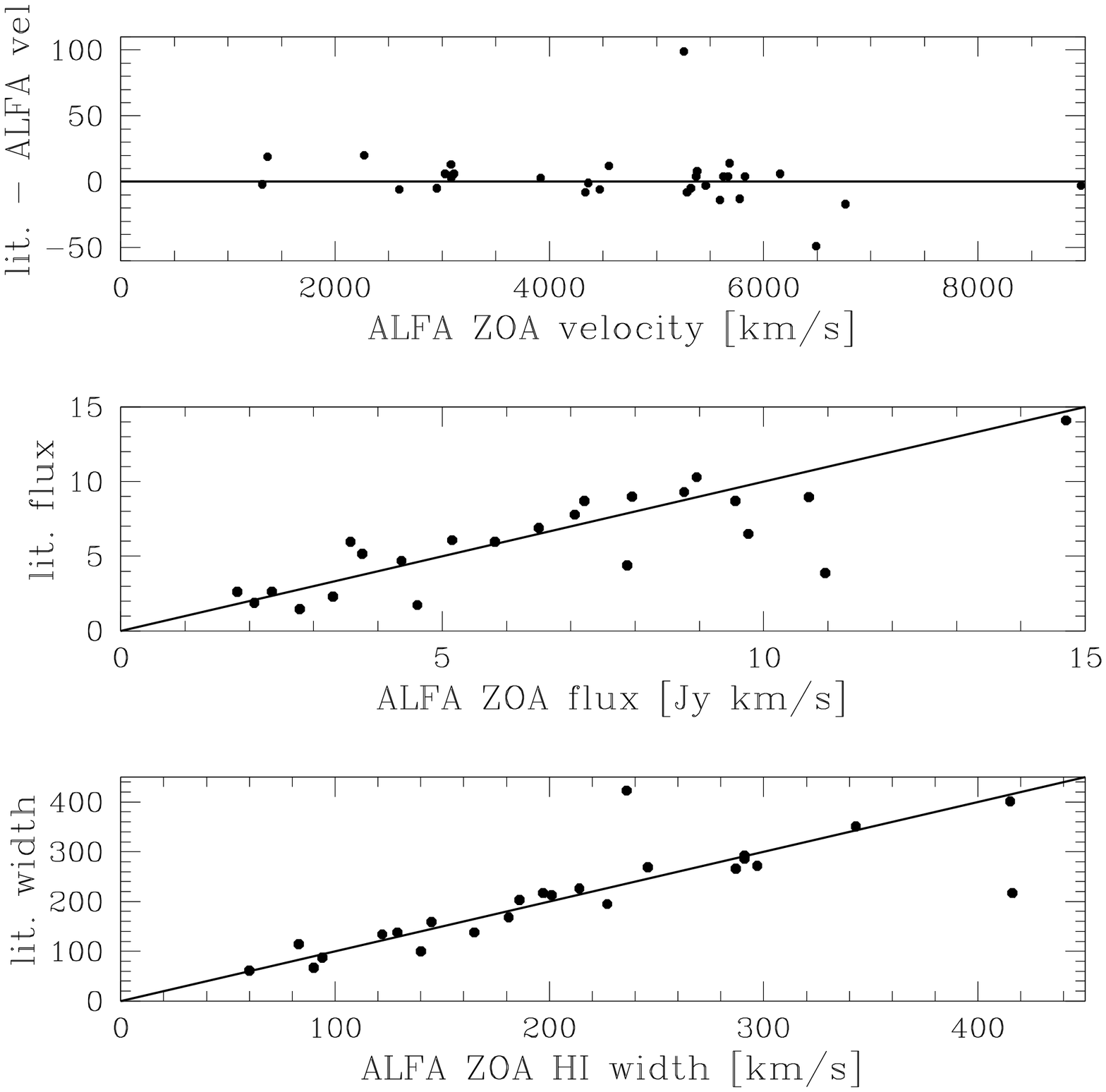}
\caption{Comparison of HI parameters of the ALFA ZOA galaxies with galaxies having 
HI measurements in the literature.
Top panel:  difference between the literature systemic velocity and the ALFA ZOA systemic
velocity measurement versus the ALFA ZOA measurement, with line of zero velocity difference indicated.
Middle panel:  flux density comparison, with line of slope one shown.  Bottom panel:  
linewidth at 50\% peak flux comparison, with line of slope one shown.
The most outlying point in each plot refers to ADBS051733+1934 = ALFAZOA J0517+1936.
\label{fig6}}
\end{figure}

\subsection{Inner Galaxy Detections}

The first region observed, toward the heavily-obscured inner Galaxy,
was selected to overlap the northern extension of the Parkes ZOA
survey (Donley \etal 2005) with similar sensitivity, so we could check
performance of the observing and analysis systems with known HI
sources.  We detected 10 HI galaxies in this area, including 7 of the
Parkes galaxies, and three more HI sources: one associated with an IR
galaxy, and two newly-discovered galaxies.  Because of the thick
obscuration in the optical and confusion in the NIR in this region,
only one ALFA ZOA galaxy (10\%) has a cataloged counterpart in any
other waveband.  This region was selected because it had already been
covered by an HI survey of similar depth, thus we report no newly
uncovered large-scale structures here, but refer the reader to Donley
\etal 2005 for plots of HI galaxy distribution in this region, and
connection to high latitude large-scale structure.

\subsection{Outer Galaxy Detections}

In the lower-extinction, less confused outer Galaxy region, 49 of the
62 galaxies (79\%) detected by ALFA ZOA have counterparts (mostly
2MASS) but only 26 (42\%) have previously published redshifts.  The
sky distribution is shown

\clearpage
\begin{deluxetable}{rrrrrrrrrrrrr}
\tabletypesize{\tiny}
\rotate
\tablecolumns{13}
\tablewidth{0pc}
\tablecaption{Multiwavelength Counterparts in the Literature of ALFA ZOA Precursor Detections}
\tablehead{\colhead{ALFAZOA} & \colhead{b} & \colhead{A$_{\rm B}$} & \colhead{HI} & \colhead{sep} & \colhead{$\Delta V$} 
& \colhead{Optical} 
& \colhead{sep} &\colhead{2MASS} & \colhead{sep} & \colhead{IRAS} & \colhead{sep} & 
\colhead{$\Delta V$} \\

\colhead{} & \colhead{(deg)} & \colhead{(mag)} & \colhead{} & \colhead{(\arcmin)} & \colhead{(\kms)} & \colhead{} &
\colhead{(\arcmin)} & \colhead{} & \colhead{(\arcmin)} & \colhead{} & \colhead{(\arcmin)} & \colhead{(\kms)}} 

\startdata
J1855+0737 & 2.63 & 17.1 & HIZOAJ1855+07 & 2.6 & 6 & \nodata & \nodata & \nodata & \nodata & \nodata & \nodata & \nodata \\
J1901+0651 & 0.88 & 23.8 & HIZOAJ1901+06 & 1.2 & -5 &\nodata  &\nodata  &\nodata  &\nodata  &\nodata  &\nodata  &\nodata  \\
&  &  & aka HIPASSJ1901+06 &  &  &  &  &  &  &  &  &  \\
J1906+0734 & 0.09 & 66.5 & HIZOAJ1906+07 & 3.1 & 13 &\nodata  &\nodata  &\nodata  &\nodata  &\nodata  &\nodata  &\nodata  \\
J1908+0559 & -1.03 & 28.3 & HIZOAJ1908+05 & 0.2 & 12 &\nodata  &\nodata  &\nodata  &\nodata  &\nodata  &\nodata  &\nodata  \\
J1917+0749 & -2.16 & 8.4 & HIZOAJ1917+07 & 0.8 & 6 &\nodata  &\nodata  &\nodata  &\nodata  &\nodata  &\nodata  &\nodata  \\
J1920+0610 & -3.64 & 4.5 & IRAS19182+0604 & 0.4 & -49 & ZOAG041.78-03.65 & 0.4 & 2M19204508+0610446 & 1.1 & IRAS19182+0604 & 0.4 &\nodata  \\
J1922+0818 & -2.94 & 4.4 & HIZOAJ1921+08 & 1.5 & 6 &\nodata  &\nodata  &\nodata  &\nodata  &\nodata  &\nodata  &\nodata  \\
&  &  & aka HIPASSJ1922+08 &  &  &  &  &  &  &  &  &  \\
J1925+0816 & -3.81 & 3.6 & HIZOAJ1926+08 & 1.5 & 3 &\nodata  &\nodata  &\nodata  &\nodata  &\nodata  &\nodata  &\nodata  \\
J0449+2141 & -14.49 & 1.7 & HIPASSJ0449+21 & 0.2 & 3 & UGC03177 & 0.2 &\nodata  &\nodata  &\nodata  &\nodata  &\nodata  \\
J0457+2241 & -12.38 & 1.8 &\nodata  &\nodata  &\nodata  &\nodata  &\nodata  & 2M04575576+2239592 & 2.1 &\nodata  &\nodata  &\nodata  \\
& \nodata & \nodata & \nodata & \nodata & \nodata & \nodata & \nodata & 2M04580046+2240042 & 2.6 & \nodata & \nodata & \nodata \\
J0459+2102 & -12.98 & 1.6 &\nodata  &\nodata  &\nodata  &\nodata  &\nodata  & 2M04595592+2102028 & 0.5 &\nodata  &\nodata  &\nodata  \\
& \nodata & \nodata &\nodata  & \nodata & \nodata & \nodata & \nodata & 2M04595870+2058438 & 3.4  &\nodata  &\nodata  &\nodata  \\
J0505+2059 & -12.02 & 2.4 & UGC03243 & 0.2 & -5 & UGC03243 & 0.2 & 2M05051205+2059000 & 0.2 & \nodata & \nodata & \nodata \\
J0509+1930 & -12.09 & 2.6 & \nodata & \nodata & \nodata & \nodata & \nodata & 2M05091428+1930279 & 0.2 & \nodata & \nodata & \nodata \\
J0509+1949 & -11.83 & 2.6 & CAP0506+19a & 0.1 & 4 & \nodata & \nodata & 2M05094137+1949105 & 0.1 & \nodata & \nodata & \nodata \\
& \nodata & \nodata & CAP0506+19b & 1.3 & \nodata & \nodata & \nodata & 2M05094420+1948005 & 1.3 &  \nodata &  \nodata &  \nodata \\ 
J0510+2138 & -10.66 & 2.5 & \nodata & \nodata & \nodata & \nodata & \nodata & 2M05102162+2139250 & 0.8 & \nodata & \nodata & \nodata \\
J0510+2044 & -11.10 & 3.0 & \nodata & \nodata & \nodata & \nodata & \nodata & 2M05104468+2044540 & 0.2 & \nodata & \nodata & \nodata \\
& \nodata & \nodata & \nodata & \nodata & \nodata & \nodata & \nodata & 2M05104097+2042320 & 2.3 & \nodata & \nodata & \nodata \\
J0512+2039 & -10.88 & 3.0 & \nodata & \nodata & \nodata & \nodata & \nodata & 2M05120483+2035593 & 3.4 & \nodata & \nodata & \nodata \\
J0512+2210 & -9.91 & 2.1 & \nodata & \nodata & \nodata & \nodata & \nodata & 2M05124109+2209486 & 0.4 & \nodata & \nodata & \nodata \\
& \nodata  & \nodata  & \nodata  & \nodata  & \nodata  & \nodata  & \nodata  & 2M05124879+2210306 & 1.8 & \nodata  & \nodata  & \nodata  \\
J0513+2033 & -10.70 & 2.9 & CAP0510+20 & 0.4 & -14 & \nodata & \nodata & 2M05132510+2033254 & 0.4 & IRAS05104+2029 & 0.4 & -67 \\
J0514+2031 & -10.54 & 2.8 & \nodata & \nodata & \nodata & \nodata & \nodata & 2M05141577+2032095 & 0.4 & \nodata & \nodata & \nodata \\
J0515+1928 & -10.94 & 2.5 & \nodata & \nodata & \nodata & \nodata & \nodata & 2M05151475+1931285 & 3.1 & \nodata & \nodata & \nodata \\
J0515+2209 & -9.41 & 2.3 & CAP0512+22 & 1.7 & -3 & \nodata & \nodata & 2M05151674+2210415 & 1.7 & IRAS05122+2207 & 1.7 & \nodata \\
& \nodata  & \nodata  & \nodata  & \nodata  & \nodata  & \nodata  & \nodata  & 2M05151487+2209455 & 1.9 & \nodata  & \nodata  & \nodata  \\
& \nodata  & \nodata  & \nodata  & \nodata  & \nodata  & \nodata  & \nodata  & 2M05151551+2211245 & 2.4 & \nodata  & \nodata  & \nodata  \\
& \nodata  & \nodata  & \nodata  & \nodata  & \nodata  & \nodata  & \nodata  & 2M05152921+2212137 & 2.9 & \nodata  & \nodata  & \nodata  \\
J0515+1921 & -10.92 & 2.5 & CAP0512+19 & 0.1 & -8 & PGC016993 & 0.1 & 2M05154347+1921467 & 0.1 & IRAS05127+1918 & 0.1 & 48 \\
& \nodata  & \nodata  & \nodata  & \nodata  & \nodata  & \nodata  & \nodata  & 2M05153795+1919117 & 2.8 & \nodata  & \nodata  & \nodata  \\
J0516+2051 & -9.87 & 2.6 & CAP0513+20 & 0.4 & -17 & \nodata & \nodata & 2M05164511+2052126 & 0.4 & IRAS05137+2049 & 0.4 & \nodata \\
J0517+2120 & -9.52 & 2.7 & \nodata & \nodata & \nodata & \nodata & \nodata & 2M05170539+2121519 & 2.3 & \nodata & \nodata & \nodata \\
J0517+1936 & -10.40 & 2.5 & ADBS051733+1934 & 2.4 & 99 & \nodata & \nodata & 2M05174145+1936010 & 0.2 & \nodata & \nodata & \nodata \\
J0518+1909* & -10.40 & 2.2 & UGC03285 & 0.8 & 7 & PGC017079 & 0.8 & 2M05185631+1910219 & 0.8 & \nodata & \nodata & \nodata \\
J0519+2256* & -8.15 & 3.6 & HIPASSJ0519+22 & 1.4 & 2 & \nodata & \nodata & 2M05193869+2257031 & 0.4 & \nodata & \nodata & \nodata \\
& \nodata  & \nodata  & \nodata  & \nodata  & \nodata  & \nodata  & \nodata  & 2M05193160+2258441 & 2.6 & \nodata & \nodata & \nodata \\
J0521+1923 & -9.72 & 1.9 & 2M05214377+1923370 & 0.5 & -1 & \nodata & \nodata & 2M05214377+1923370 & 0.5 & \nodata & \nodata & \nodata \\
J0524+2129 & -8.13 & 3.1 & \nodata & \nodata & \nodata & \nodata & \nodata & 2M05235899+2128531 & 0.3 & IRAS05209+2126 & 0.3 & 14 \\
& \nodata  & \nodata  & \nodata  & \nodata  & \nodata  & \nodata  & \nodata  & 2M05240799+2131455 & 3.3 & \nodata  & \nodata  & \nodata  \\
J0525+2151 & -7.64 & 2.6 & CAP0522+21 & 0.3 & 4 & UGC03304 & 0.3 & 2M05252929+2151220 & 0.3 & IRAS05224+2148 & 0.3 & \nodata \\
& \nodata  & \nodata  & \nodata  & \nodata  & \nodata  & \nodata  & \nodata  & 2M05252856+2147520 & 3.5 & \nodata  & \nodata  & \nodata  \\
J0526+1957 & -8.46 & 3.0 & \nodata & \nodata & \nodata & \nodata & \nodata & 2M05263532+1957471 & 0.3 & \nodata & \nodata & \nodata \\
J0545+1925 & -5.03 & 2.6 & \nodata & \nodata & \nodata & \nodata & \nodata & 2M05451023+1925174 & 0.4 & \nodata & \nodata & \nodata \\
J0548+1911* & -4.39 & 3.4 & ZOAG188.73-04.39 & 0.3 & -12 & ZOAG188.73-04.39 & 0.3 & 2M05485392+1911467 & 0.3 & IRAS05459+1910 & 0.3 & \nodata \\
J0552+2255* & -1.71 & 5.9 & \nodata & \nodata & \nodata & \nodata & \nodata & 2M05525788+2255555 & 2.5 & \nodata & \nodata & \nodata \\
J0559+2109 & -1.21 & 6.3 & \nodata & \nodata & \nodata & \nodata & \nodata & 2M05594285+2108465 & 0.6 & \nodata & \nodata & \nodata \\
J0602+2204 & -0.25 & 5.0 & \nodata & \nodata & \nodata & \nodata & \nodata & 2M06021042+2203017 & 1.3 & \nodata & \nodata & \nodata \\
J0602+2201 & -0.20 & 4.7 & \nodata & \nodata & \nodata & ZOAG187.89-00.20 & 0.3 & 2M06023546+2201525 & 0.3 & IRAS05595+2201 & 0.3 & -6 \\
J0604+2201 & 0.27 & 4.5 & \nodata & \nodata & \nodata & \nodata & \nodata & 2M06045592+2201120 & 0.3 & IRAS06019+2201 & 0.3 & \nodata \\
J0605+1927 & -0.88 & 6.6 & HIPASSJ0605+19 & 2.3 & -14 & \nodata & \nodata & 2M06052653+1927297 & 0.4 & \nodata & \nodata & \nodata \\
& \nodata & \nodata & \nodata & \nodata & \nodata & \nodata & \nodata & 2M06051648+1927519 & 2.8 & \nodata & \nodata & \nodata \\
J0606+2036 & -0.05 & 6.4 & \nodata & \nodata & \nodata & \nodata & \nodata & 2M06063438+2037119 & 2.8 & \nodata & \nodata & \nodata \\
& \nodata & \nodata & \nodata & \nodata & \nodata & \nodata & \nodata & 2M06064755+2039219 & 3.0 & \nodata & \nodata & \nodata \\
J0607+2152 & 0.62 & 3.4 & IRAS06040+2152 & 0.1 & 4 & ZOAG188.54+00.62 & 0.1 & 2M06070237+2152194 & 0.1 & IRAS06040+2152 & 0.1 & \nodata \\
J0612+2113 & 1.47 & 3.8 & \nodata & \nodata & \nodata & \nodata & \nodata & 2M06124440+2113302 & 0.3 & \nodata & \nodata & \nodata \\
& \nodata & \nodata & \nodata & \nodata & \nodata & \nodata & \nodata & 2M06124197+2113382 & 0.6 & \nodata & \nodata & \nodata \\
& \nodata & \nodata & \nodata & \nodata & \nodata & \nodata & \nodata & 2M06125062+2113052 & 1.5 & \nodata & \nodata & \nodata \\
& \nodata & \nodata & \nodata & \nodata & \nodata & \nodata & \nodata & 2M06122966+2114228 & 3.5 & \nodata & \nodata & \nodata \\
J0614+1934 & 1.05 & 6.0 & \nodata & \nodata & \nodata & \nodata & \nodata & 2M06143273+1934449 & 0.5 & \nodata & \nodata & \nodata \\
& \nodata & \nodata & \nodata & \nodata & \nodata & \nodata & \nodata & 2M06143467+1932174 & 2.8 & \nodata & \nodata & \nodata \\
J0616+2142 & 2.41 & 4.9 & \nodata & \nodata & \nodata & \nodata & \nodata & 2M06161256+2142233 & 0.1 & \nodata & \nodata & \nodata \\
J0620+2008 & 2.63 & 3.5 & ADBS062054+2008 & 0.4 & -2 & \nodata & \nodata & 2M06205100+2006513 & 1.7 & \nodata & \nodata & \nodata \\
&  &  & aka HIPASSJ0620+20 &  &  &  &  &  &  &  &  &  \\
J0621+2010 & 2.69 & 3.6 & ADBS062103+2010 & 0.7 & 20 & \nodata & \nodata & \nodata & \nodata & \nodata & \nodata & \nodata \\
J0621+2256* & 4.00 & 6.6 & \nodata & \nodata & \nodata & \nodata & \nodata & 2M06210911+2255353 & 0.4 & \nodata & \nodata & \nodata \\
J0621+2142 & 3.56 & 2.7 & \nodata & \nodata & \nodata & \nodata & \nodata & 2M06214885+2142406 & 0.3 & \nodata & \nodata & \nodata \\
J0629+2151 & 5.24 & 1.4 & \nodata & \nodata & \nodata & \nodata & \nodata & 2M06294499+2150344 & 2.1 & \nodata & \nodata & \nodata \\
J0633+2102 & 5.68 & 1.4 & HIPASSJ0633+21 & 0.3 & -3 & UGC03489 & 0.3 & 2M06333350+2102135 & 0.3 & IRAS06305+2104 & 0.3 & \nodata \\
& \nodata & \nodata & \nodata & \nodata & \nodata & \nodata & \nodata & 2M06334119+2101295 & 1.6 & \nodata & \nodata & \nodata \\
J0635+2039 & 5.86 & 1.4 & HIPASSJ0635+20 & 0.4 & -8 & \nodata & \nodata & 2M06351132+2039470 & 0.4 & \nodata & \nodata & \nodata \\
J0638+2238 & 7.33 & 0.7 & HIPASSJ0637+22 & 0.4 & 19 & UGC03503 & 0.4 & \nodata & \nodata & \nodata & \nodata & \nodata \\
J0639+2044 & 6.81 & 0.9 & UGC03505 & 0.3 & 4 & UGC03505 & 0.3 & 2M06393660+2044335 & 0.3 & \nodata & \nodata & \nodata \\
J0642+2108 & 7.50 & 0.7 & CAP0639+21 & 0.3 & 8 & \nodata & \nodata & 2M06420364+2108345 & 0.3 & IRAS06391+2111 & 0.3 & \nodata \\
& \nodata & \nodata & \nodata & \nodata & \nodata & \nodata & \nodata & 2M06420114+2108435 & 0.6 & \nodata & \nodata & \nodata \\
J0643+2252* & 8.49 & 1.2 & UGC03516 & 0.6 & 2 & PGC019483 & 0.6 & \nodata & \nodata & \nodata & \nodata & \nodata \\
J0645+2225 & 8.82 & 1.3 & HIPASSJ0645+22 & 0.2 & -6 & UGC03534 & 0.2 & 2M06454136+2225464 & 0.2 & \nodata & \nodata & \nodata \\
& \nodata & \nodata & \nodata & \nodata & \nodata & \nodata & \nodata & 2M06454263+2224583 & 0.7 & \nodata & \nodata & \nodata \\
\enddata
\end{deluxetable}

\clearpage

\noindent in the right panel of Figure 1, which also indicates the
location of the Crab nebula.  This strong continuum source locally
raised the system temperature and disturbed the spectral baselines.
The noise in the underdense region to the Galactic North of the Crab
nebula is comparable to the average survey noise, indicating that the
underdensity in galaxies there is probably real.

The velocity distribution of the ALFA ZOA galaxies plotted as a wedge
diagram in RA, collapsed around the central declination of the search
($21^\circ$), is shown in Figure 7.  For this plot, cataloged galaxies
with known velocity are also shown, and the wedge continues in RA
beyond the limits of our survey (which is shown by the central wedge).
Most notable in the ALFA ZOA galaxy distribution is an apparent
overdensity of galaxies near $5^{\rm h}$, and between 5000 - 6000 \kms
($\ell \sim 183^\circ, b \sim -10^\circ$).  This overdensity is also
evident in the optical and IR-selected sample of Pantoja \etal (1997,
2000), and the 2MASS sample of van Driel \etal (2009).  The most
obvious concentration near this velocity seen in the cataloged
galaxies is the Cancer cluster, with its finger of God apparent at
$8^{\rm h} 20^{\rm m}$, 4500 \kms.  The Perseus-Pisces chain, at
similar velocity, enters the ZOA at $\ell \sim 160^\circ$ (apparent in
Fig 1).  The ALFA ZOA concentration lies very roughly between the two
on the sky.  This overdensity lies close to the Orion concentration
evident in the 2MASS Redshift Survey density field reconstruction
(Erdogdu \etal 2006) in the shell at 6000 \kms.  Also nearby in the
reconstruction is the feature ``C5'', at $\ell \sim 195^\circ, b \sim
0^\circ$, with velocity peak at 5000 \kms.  This may be evidence of a
real overdensity of galaxies consistent with the large-scale structure
reconstruction at low-Galactic latitude.

Also apparent in the wedge plot is the Taurus void (center at RA =
3.5$^{\rm h}$, Dec = +20$^\circ$, v = 4000 km s$^{-1}$) which appears
as underdensity on the low RA side, and possibly a portion of the
Gemini void (center at RA = 6$^{\rm h}$, Dec = +40$^\circ$, v = 3000
km s$^{-1}$).

\begin{figure}[t]
\plotone{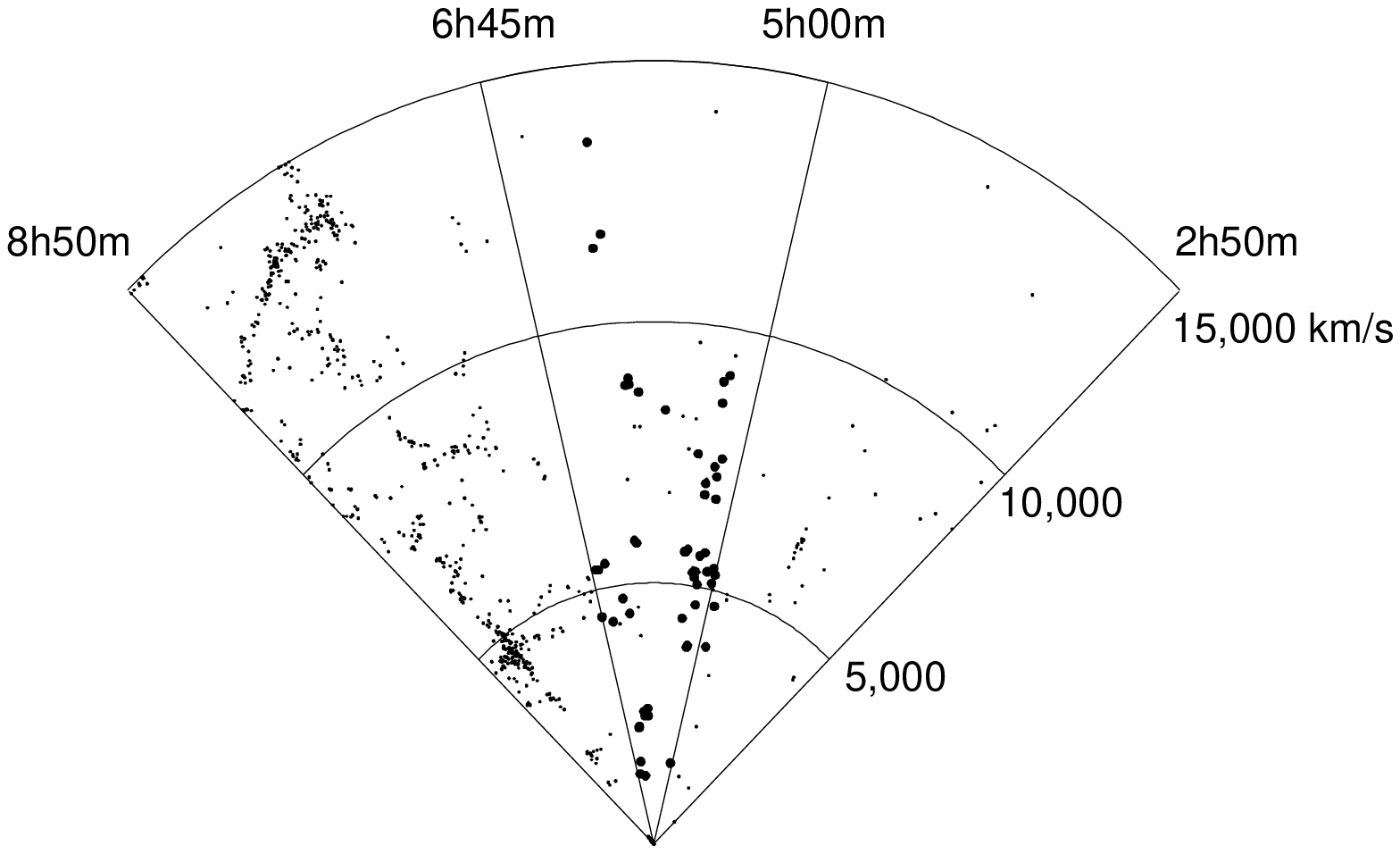}
\epsscale{1.2}
\caption{Distribution of ALFA ZOA outer Galaxy detections (heavy dots) within the search
area (inner wedge) and LEDA galaxies (small dots).  The wedge is collapsed over the declination
range of the ALFA ZOA detections, a 3 degree range centered on declination $21^\circ$.
The RA range is extended to show known distribution in LEDA galaxies further from the Galactic
plane.  The ALFA ZOA detections shown
outside of the inner wedge were found beyond the full-sensitivity search region, where scans 
did not overlap, but still had some search data.
\label{fig7}}
\end{figure}

\section{The Future:  Surveying the Entire Arecibo ZOA}

The precursor observations of a relatively small area of the ZOA visible to the Arecibo
Radio Telescope have
shown that we can detect new galaxies behind the Milky Way, and with the redshifts
provided by the 21-cm spectral line of HI, delineate large-scale structure at 
low Galactic latitude.
These observations are only a prelude to a large survey of the ZOA, to
be conducted in two phases, one shallow, and one deep.
A shallow survey of the entire inner Galaxy region, $\ell = 30^\circ - 75^\circ,
|b| \le 10^\circ, \sim1000$ square degrees, 
began in May 2008, in the nodding scan mode described here.
Extrapolating from the detection rate of the precursor observations,
we anticipate a final catalog in this region of about 500 galaxies.
The vast majority will be new detections, 
as only 1/10th of the  detections from the precursor inner Galaxy observations have a 2MASS
counterpart in the extended source catalog.
This is not surprising due to the high stellar surface density in the Galactic first
quadrant.
For the galaxies which do have counterparts at other wavelengths, 
we will provide high-quality redshifts.
This area of the sky is almost uncharted territory (Fig 1), so we anticipate the discovery
of many unknown galaxies and structures.
Observations for this phase of the survey were completed in 2009.

Commensal observations with a deep survey for pulsars within $|b| \le 5^\circ$ 
began in July 2009.
The multi-year survey began in the inner Galaxy, with intention to
extend the survey to the outer Galaxy ($\ell = 170^\circ - 215^\circ$) in the future.
These sensitive data will also be used by a team interested in surveying Galactic radio recombination lines.
The integration time planned for this project is a lengthy 268 seconds per pointing 
(compare to 8 sec per beam in the nodding scan mode), 
with the sky covered by tiles of individual pointings.
The higher sensitivity of this deep phase of the survey is expected to yield thousands of
galaxies, including local low-mass objects, and HI galaxies at large distances.

This work is based on observations done at Arecibo Observatory.
The Arecibo Observatory is part of the National Astronomy and
Ionosphere Center, which is operated by Cornell University
under a cooperative agreement with the National Science Foundation.
We thank the Arecibo Observatory staff for help organizing these
commensal observations, and Mark Calabretta for quickly adjusting
the Parkes multibeam reduction software to accept the new format.

This research has made use of the NASA/IPAC Extragalactic Database (NED)
which is operated by the Jet Propulsion Laboratory, California Institute
of Technology, under
contract with the National Aeronautics and Space Administration.
We acknowledge the usage of the HyperLeda database (http://leda.univ-lyon1.fr).

PAH gratefully acknowledges support from NSF grant AST-0506676.


\begin{thebibliography}{}

\bibitem[]{}
Auld, R., Minchin, R. F., Davies, J. I., Catinella, B., van Driel, W.,
Henning, P. A., Linder, S., Momjian, E., Muller, E., O'Neil, K.,
Sabatini, S., Schneider, S., Bothun, G., Cortese, L., Disney, M.,
Hoffman, G. L., Putman, M., Rosenberg, J. L., Baes, M., de Blok, W. J.
G., Boselli, A., Brinks, E., Brosch, N., Irwin, J., Karachentsev, I. D.,
Kilborn, V. A., Koribalski, B., \& Spekkens, K. ~2006, \mnras, 371, 1617

\bibitem[]{}
Barnes, D.G., Staveley-Smith, L., de Blok, W.J.G., Oosterloo, T., Stewart, I.M., 
Wright, A.E., Banks, G.D., Bhathal, R., Boyce, P.J., Calabretta, M.R.,
Disney, M.J., Drinkwater, M.J., Ekers, R.D.,
Freeman, K.C., Gibson, B.K., Green, A.J., Haynes, R.F.,
te Lintel Hekkert, P., Henning, P.A., Jerjen, H., Juraszek, S.,
Kesteven, M.J., Kilborn, V.A., Knezek, P.M., Koribalski, B.,
Kraan-Korteweg, R.C., Malin, D.F., Marquarding, M., Minchin, R.F.,
Mould, J.R., Price, R.M., Putman, M.E., Ryder, S.D., Sadler, E.M.,
Schr\"oder, A., Stootman, F., Webster, R.L., Wilson, W.E., \&
Ye, T. ~2001, \mnras, 322, 486

\bibitem[]{}
Branchini, E., 
Teodoro, L., Frenk, C. S., Schmoldt, I., Efstathiou, G., White, S. D. M., 
Saunders, W., Sutherland, W., Rowan-Robinson, M., Keeble, O., Tadros, H.,
Maddox, S., \& Oliver, S. ~1999, \mnras ~308, 1

\bibitem[]{}
Cortese, L., Minchin, R.F., Auld, R.R., Davies, J.I., Catinella, B., Momjian, E., 
Rosenberg, J.L., Taylor, R., Gavazzi, G., O'Neil, K., Baes, M., Boselli, A., 
Bothun, G., Koribalski, B., Schneider, S., \& van Driel, W. 2008, \mnras, 383, 1519

\bibitem[]{}
Donley, J.L., Staveley-Smith, L., Kraan-Korteweg, R.C., Islas-Islas, J.M., Schr\"oder, A.,
Henning, P.A., Koribalski, B., Mader, S., \& Stewart, I.  ~2005, \aj ~129, 220

\bibitem[]{}
Erdogdu, P., Lahav, O., Huchra, J.P., Colless, M., Cutri, R.M., Falco, E., George, T., Jarrett, T., Jones, D.H., Macri, L.M.,
Mader, J., Martimbeau, N., Pahre, M.A., Parker, Q.A., Rassat, A., \& Saunders, W. ~2006, MNRAS, 373, 45

\bibitem[]{}
Fairall, A.P. 1998, Large-Scale Structures in the Local Universe, 
(Chichester: Wiley)

\bibitem[]{}
Freudling, W. Staveley-Smith, L., Calabretta, M., Catinella, B., van Driel, W., Linder, S., Minchin, R., 
Momjian, E., Zwaan, M., \& AUDS Team ~2005, BAAS 37, 1316

\bibitem []{}
Gooch, R.E., 1996, in ASP Conf.~Ser.~101, Astronomical Data Analysis Software and Systems V, 
ed. G. H. Jacoby \& J. Barnes (San Francisco:  ASP), 80

\bibitem[]{}
Giovanelli, R.,  
Haynes, M.P., Kent, B.R., Perillat, P., Saintonge, A., Brosch, N., 
Catinella, B., , Hoffman, G.L., Stierwalt, S., Spekkens, K., Lerner, M.L., 
Masters, K.L., Momjian, E., Rosenberg, J.L., Springob, C.M., 
Boselli, A., Charmandaris, V., Darling, J.K., Davies, J.,
Lambas, D.G., Gavazzi, G., Giovanardi, C., Hardy, E.,
Hunt, L.K., Iovino, A., Karachentsev, I.D., Karachentseva, V.E.,
Koopmann, R.A., Marinoni, C., Minchin, R., Muller, E., 
Putman, M., Pantoja, C., Salzer, J.J., Scodeggio, M.,
Skillman, E., Solanes, J.M., Valotto, C., 
van Driel, W., \& van Zee, L. ~2005, \aj ~130, 2598

\bibitem[]{}
Henning, P.A., Kraan-Korteweg, R.C., Rivers, A.J., Loan, A.J., Lahav, O., 
\& Burton, W.B. 1998, \aj ~115, 584

\bibitem[]{}
Henning, P.A., Staveley-Smith, L., Ekers, R.D., Green, A.J., Haynes, R.F.,
Juraszek, S., Kesteven, M.J., Koribalski, B.S., Kraan-Korteweg. R.C., Price, R.M.,
Sadler, E.M., and Schr\"oder, A. ~2000, \aj ~119, 2686

\bibitem[]{}
Henning, P.A., Kraan-Korteweg, R.C., \& Staveley-Smith, L. 2005, in workshop on 
``Nearby Large-Scale Structures \& the Zone of Avoidance'', ASP Conf. Ser. 329, 
eds. A.P. Fairall \& P.A. Woudt, (San Francisco: ASP), 199

\bibitem[]{}
Huchra, J., Jarrett, T., Skrutskie, M., Cutri, R., Schneider, S., Macri, L., Steining, R., 
Mader, J., Martimbeau, N., \& George, T. ~2005, in workshop on ``Nearby Large-Scale Structures 
\& the Zone of Avoidance'', ASP Conf. Ser. 329, eds. A.P. Fairall \& P.A. Woudt, 
(San Francisco: ASP), 135 

\bibitem[]{}
Karachentsev, I.D.,
Sharina, M.E., Makarov, D.I., Dolphin, A.E., Grebel, E.K.,  Geisler, D.,  Guhathakurta, P., Hodge, P.W., Karachentseva, V.E., 
Sarajedini, A., \& Seitzer, P. ~2002, \aap ~389, 812

\bibitem[]{}
Kraan-Korteweg, R.C. 1986, \aaps ~66, 255

\bibitem[]{}
Kraan-Korteweg, R.C., Koribalski, B.S., \& Juraszek, S. 1999, in ESO/ATNF 
Workshop on ``Looking Deep in the Southern Sky'', eds. R. Morganti 
\& Couch, Springer, 23

\bibitem[]{}
Kraan-Korteweg, R.C., Staveley-Smith, L., Donley, J., Koribalski, B. 
\& Henning, P.A. 2005, in IAU Symp.~216, Maps of the Cosmos, 
eds.~M.~Colless and L.~Staveley-Smith,
(ASP: San Francisco), 203

\bibitem[]{}
Lamm, R., Minchin, R., Momjian, E., Lowenthal, J., Henning, P., Catinella, B., \&  ALFA ZOA Collaboration
2007, BAAS, 39, 965

\bibitem[]{}
Lu, N.Y., Dow, M.W., Houck, J.R., Salpeter, E.E., \& Lewis, B.M. 1990, \apj, 357, 388

\bibitem[]{}
Pantoja, C.A., Altschuler, D.R., Giovanardi, C., \& Giovanelli,
R. 1997, AJ, 113, 905

\bibitem[]{}
Pantoja, C.A., Giovanardi, C., Altschuler, D.R., Huchra, J.P., \& Giovanelli, R. 2000,
ASP Conf. Ser. 218,
eds. R.C.~Kraan-Korteweg, P.A.~Henning, \& H.~Andernach, (San Francisco: ASP), 219

\bibitem[]{}
Peebles, P.J.E., Phelps, S.D., Shaya, E.J., \& Tully, R.B. 2001, \apj, ~554, 104

\bibitem[]{}
Rivers, A.J. 2000, Ph.D. thesis, Univ. of New Mexico

\bibitem[]{}
Rosenberg, J.L. \& Schneider, S.E. 2000, \apjs, 130, 177 

\bibitem[]{}
Saintonge, A. 2007, \aj, 133, 2087

\bibitem[]{}
Sault, R.J., Teuben, P.J., \& Wright, M.C.H. 1995, in ASP Conf. Ser. 77, 
Astronomical Data Analysis Software and Systems IV, ed. R.A. Shaw, H.E. Payne, 
\& J.J.E. Hayes (San Francisco: ASP), 433

\bibitem[]{}
Schlegel, D.J., Finkbeiner, D.P., \& Davis, M., 1998, \apj, ~500, 525

\bibitem[]{}
Shafi, Nebiha 2008, M.Sc. thesis, Univ. of Cape Town 

\bibitem[]{}
Skrutskie, M.F., Cutri, R. M., Stiening, R., Weinberg, M. D., Schneider, S., 
Carpenter, J. M., Beichman, C., Capps, R., Chester, T., Elias, J., 
Huchra, J.,Liebert, J., Lonsdale, C.,
Monet, D.G., Price. S., Seitzer, P., Jarrett, T.,  Kirkpatrick, J.D., Gizis, J.E.,
Howard, E., Evans. T., Fowler, J., Fullmer, L., Hurt, R., Light, R.,
Kopan, E.L.,  Marsh, K.A., McCallon, H.L., 
Tam, R., Van Dyk, S., Wheelock, S.  ~2006, \aj ~131, 1163

\bibitem[]{}
Stanimirovic, S.,  Putman, M., Heiles, C., Peek, J.E.G., Goldsmith, P.F., Koo, B.-C.,Krco, M., Lee, J.-J.,  Mock, J., 
Muller, E., Pandian, J.D., Parsons, A., Tang, Y., \& Werthimer, D. ~2006, \apj ~653, 1210

\bibitem[]{}
van Driel, W., Schneider, S.E., Kraan-Korteweg, R.C., \& Monnier Ragaigne, D. ~2009,
\aap ~505, 29

\bibitem[]{}
Wong, O.I., Ryan-Weber, E.V., Garcia-Appadoo, D.A., Webster, R.L.,
Staveley-Smith, L., Zwaan, M.A., Meyer, M.J., Barnes, D.G.,
Kilborn, V.A., Bhathal, R., de Blok W.J.G., Disney, M.J., Doyle, M.T.,
Drinkwater, M.J., Ekers, R.D., Freeman, K.C., Gibson, B.K., Gurovich, S.,
Harnett, J., Henning, P.A., Jerjen, H., Kesteven, M.J., Knezek, P.M.,
Koribalski, B.S., Mader, S., Marquarding, M., Minchin, R.F., O'Brien, J.,
Putman, M.E., Ryder, S.D., Sadler, E.M., Stevens, J., Stewart, I.M.,
Stootman, F., and Waugh, M. ~2006, \mnras ~371, 1855

\end{thebibliography}
\end{document}